\documentclass[aip,showpacs,author-year,preprint,cha,superscriptaddress]{revtex4-1}
\usepackage{hyperref}
\usepackage{graphicx}
\usepackage{amssymb}
\usepackage{colordvi}

\begin{document}

\title{Wall slip and flow of concentrated hard-sphere colloidal suspensions}

\date{\today}

\author{P.~Ballesta}
\affiliation{Scottish Universities Physics Alliance (SUPA) and School of Physics and Astronomy, The University of Edinburgh, Kings Buildings, Mayfield Road, Edinburgh EH9 3JZ, United Kingdom.}
\affiliation{IESL-FORTH, Heraklion 71110, Crete, Greece.}
\author{G.~Petekidis}
\affiliation{IESL-FORTH, Heraklion 71110, Crete, Greece.}
\affiliation{Department of Materials Science and Technology, University of Crete, Heraklion 71110, Crete, Greece.}
\author{L.~Isa}
\affiliation{Scottish Universities Physics Alliance (SUPA) and School of Physics and Astronomy, The University of Edinburgh, Kings Buildings, Mayfield Road, Edinburgh EH9 3JZ, United Kingdom.}
\affiliation{ETH Zurich, Laboratory for Surface Science and Technology, Wolfgang-Pauli-Strasse 10, Z\"urich, Switzerland.}
\author{W.~C.~K.~Poon}
\affiliation{Scottish Universities Physics Alliance (SUPA) and School of Physics and Astronomy, The University of Edinburgh, Kings Buildings, Mayfield Road, Edinburgh EH9 3JZ, United Kingdom.}
\author{R.~Besseling}
\affiliation{Scottish Universities Physics Alliance (SUPA) and School of Physics and Astronomy, The University of Edinburgh, Kings Buildings, Mayfield Road, Edinburgh EH9 3JZ, United Kingdom.}

\begin{abstract}

We present a comprehensive study of the slip and flow of concentrated colloidal suspensions using cone-plate rheometry and simultaneous confocal imaging. In the colloidal glass regime, for smooth, non-stick walls, the solid nature of the suspension causes a transition in the rheology from Herschel-Bulkley (HB) bulk flow behavior at large stress to a Bingham-like slip behavior at low stress, which is suppressed for sufficient colloid-wall attraction or colloid-scale wall roughness. Visualization shows how the slip-shear transition depends on gap size and the boundary conditions at both walls and that partial slip persist well above the yield stress. A phenomenological model, incorporating the Bingham slip law and HB bulk flow, fully accounts for the behavior. Microscopically, the Bingham law is related to a thin (sub-colloidal) lubrication layer at the wall, giving rise to a characteristic dependence of slip parameters on particle size and concentration. We relate this to the suspension's osmotic pressure and yield stress and also analyze the influence of van der Waals interaction. For the largest concentrations, we observe non-uniform flow around the yield stress, in line with recent work on bulk shear-banding of concentrated pastes. We also describe residual slip in concentrated liquid suspensions, where the vanishing yield stress causes coexistence of (weak) slip and bulk shear flow for all measured rates.
\end{abstract}


\maketitle

\section{Introduction}

Wall slip is a widespread phenomenon in the flow of various liquids. In Newtonian liquids, interest in slip has revived due to its relevance for flow in nano-porous media, microfluidic devices and along superhydrophobic surfaces [\cite{Barrat1999, Zhu2001, BocquetSoftMat07_slip, NetoRepProPhys_Newtslip}]. The so-called slip-length $l_s$, the distance to the wall at which the velocity profile
extrapolates to zero, can reach many molecular diameters, depending on
wettability, and strongly affects flow when it is comparable to the
system dimensions. Polymer slip has also received considerable
attention, e.g. [\cite{Hatzikiriakos1991,Hatzikiriakos1992,BrochardGennesLangmuir92_boundaryslip,Westover1966,Leger1997, Mhetar1998}]. Here changes in the chain relaxation dynamics near the wall govern slip: typically a transition from weak slip at small flow
rate to strong slip ($l_s$ exceeding hundreds of micrometers [\cite
{Mhetar1998}]) for large flow rate is seen, driven by chain
disentanglement near the wall. In surfactant solutions, recent work [\cite
{SalmonPRE03,BecuPRL04_bandingdynamics,Manneville2004,LettingaPRL09_WLM_slip,Becu2007}
] has shown more complex behavior, where slip and wall interactions are
coupled to the shear-banding	 in these systems.

The most prominent examples of slip in industrial and daily applications
occur in flow of complex, multi-phase fluids [\cite
{Yoshimura1988,Barnes1995,Larson1999}]. Meaningful characterization of the
bulk flow properties of these systems requires proper insight into boundary effects [\cite{BuscallJRheo2010_slip}]. Over the last
decades, many studies of slip in these systems have appeared, e.g. in
particulate (colloidal) suspensions [\cite
{Yilmazer1989,Aral1994,HartmanKok2004,Kalyon2005,Jana1995,Cohen2006,Isa2007,Soltani,Persello}
], colloidal gels [\cite{Buscall1993, Russel2000,
WallsJRheo03_slip,VaradanSolomonJRheo2003_gelflowconfocal,GibaudMannevillePRL08_slipyield,Wassenius2005}
] and emulsions and foams [\cite{Bertola2003,Princen1985,
Meeker2004a,Meeker2004b,Salmon2003,DenkovColSurfSciA05_foamfriction,Katgert2008}].
Despite this large body of work, it is challenging to gain microscopic
insight into the nature of slip and understand its dependence on material
composition, wall properties, and flow rate. Broadly speaking, slip results
from depletion of the dispersed phase near a smooth wall, giving a low
viscosity, high shear boundary layer which reduces the \textit{apparent}
bulk viscosity \footnote{
We refer here to `intrinsic' depletion under uniform stress, as opposed to
depletion of dispersed phase due to shear migration as may arise under
non-uniform stress.}. However, both the structure and the origin of this
``layer'' vary greatly between different systems and the interplay with
(non)linear bulk rheology can give rise to slip being pronounced either at
large or small flow rate.

Slip at low stress or applied shear rate can occur in dispersions where
caging, aggregation or `jamming' leads to a solid-like microstructure and
mechanical behavior. The system then exhibits a yield stress, $\sigma_y$,
below which the micro-structure remains intact but, depending on wall
interactions and roughness, \textit{apparent} flow of the material can
still be measured. This was recently studied for concentrated emulsions and other soft
particle pastes [\cite{Meeker2004a,Seth2008}] in presence of smooth walls.
There, elasto-hydrodynamic lubrication, associated with the particles'
deformability, causes a lubrication layer between the compressed packing and
the wall with velocity dependent thickness. This gives a nonlinear relation
between the slip stress $\sigma$ and the slip velocity $v_s$, but this
mechanism only occurs for very weakly or non-repulsive particle-wall
interactions [\cite{Meeker2004b,Seth2008}]. Note that nonlinear lubrication
is of strong interest to applications of many other soft solids, e.g.
hydrogels, where combined wall-network repulsion and adsorption determines
the friction properties [\cite{Gong2010}].

For suspensions of `hard' particles, most studies of slip have been performed
for \textit{non-Brownian} systems, i.e. at large P\'{e}clet number $\mathrm{
Pe}= \dot\gamma \tau_B >> 100$, where $\dot\gamma$ is the true bulk flow
rate and $\tau_B$ the Brownian relaxation time. There, slip occurs both in solid- [\cite
{Yilmazer1989, Kalyon2005}] and liquid-like [\cite{Jana1995}] suspensions
(albeit without consensus on the quantitative behavior), but little attention has been given to the effect of
Brownian motion on slip in colloidal systems. \cite{HartmanKok2004} found
that Brownian motion prevents depletion at low shear rates in dilute
colloids, so that slip becomes apparent only for $\mathrm{Pe} \gtrsim 10$.
However, for more concentrated colloids, `crowding' competes with Brownian
relaxation. This causes structural arrest, a glass transition [\cite
{Pusey1986,Megen1998}], which for hard-sphere (HS) colloids occurs at a
volume fraction of $\phi_g \simeq 0.58$. The associated change from liquid
to solid-like rheology [Petekidis et al. (2004)] can significantly
affect the slip response.

The slip and flow behavior in these concentrated HS colloidal suspensions is
the subject of this paper. In [\cite{Ballesta2008}] we have given a short
account of some of the results, here we present a more extensive study of
the behavior. Using rheo-microscopy [\cite{Besseling2009}], we show that for 
$\phi> \phi_g$, slip becomes dominant in the rheology for smooth, non-stick
walls. We address the dependence of slip on colloid-wall (van der Waals)
interactions and concentration and quantitatively describe the transition
from slip to yielding for different combinations of confining walls via a
phenomenological model. Below the yield stress, slip causes full plug flow,
qualitatively similar to jammed emulsions and microgels [\cite{Meeker2004a,
Seth2008}], but quantitatively different; above a threshold stress, the slip
stress increases linearly with the slip velocity (Bingham slip response) due
to a sub-colloidal solvent layer with velocity-independent thickness. The
concentration dependence of the Bingham slip parameters shows a direct
relation to the osmotic pressure and yield stress of the suspensions, which
can be understood on a semi-quantitative basis but still lacks a full
theoretical description. We further show that deep in the glass regime,
shear localization effects [\cite{Besseling2010}] accompany the slip-shear
transition and discuss residual slip in concentrated liquid suspensions ($
\phi< \phi_g$), a feature not uncovered previously~[\cite{Ballesta2008}] due
to its limited effect on the bulk rheology.

The paper is structured as follows. After a description of the measurement
setup and suspensions in Sec.~\ref{sec_methods}, we present rheology and
local velocimetry results for various hard sphere suspensions, colloid-wall interactions
and confining walls conditions in Sec.~\ref{sec_results}. In Sec.~\ref
{Sec_model}, we describe the phenomenological model for slip and yielding,
generalizing the results in [\cite{Ballesta2008}] to account for geometries
with various confining walls, and in Sec.~\ref{sec_compare} we compare the
predictions with the experimental results. The physical origin of the
Bingham slip parameters is analyzed in Sec.~\ref{Sec_slipparam}, along with
the effect of van der Waals interactions. Sect.~\ref{sec_localization} and 
\ref{section_slip_liq} deal with shear localization for $\phi > \phi_g$ and
residual slip in liquid suspensions for $\phi<\phi_g$, respectively, and we conclude in Sect.~\ref{disc_concl}.

\section{Samples and methods}

\label{sec_methods}

\subsection{Colloidal suspensions}

\label{subsec_colloids}

We used polymethylmethacrylate (PMMA) colloids of various sizes (radii $
a=138 $~nm, $150$~nm, $302$~nm and a fluorescent batch with $a=652$~nm,
measured by light scattering, polydispersity $\sim 15\%$), sterically
stabilized with a poly-12-hydroxystearic acid (PHSA) layer [\cite{Barret74}] and dispersed in a refractive index (RI) matching solvent mixture of decalin and tetralin (viscosity $\eta_s=2.2$~mPas). A few measurements were
also conducted using pure decalin as solvent. Non-fluorescent (RI-matched) samples were
seeded with $\sim 0.5\%$ of the fluorescent particles (labeled with
nitrobenzoxadiazole, NBD), which served as tracers during confocal imaging
(Fig.~\ref{fig_setup}(b)). In these solvents the colloids interact almost
like perfect HSs [\cite{Bryant2002}]. In the decalin-tetralin mixture the refractive index of the colloids is $\sim 1.50-1.51$, somewhat larger than the bulk PMMA value ($1.49$), due to solvent absorbtion of the colloids and slight swelling, estimated to be $\sim 10~\%$.

Batches of different volume fractions $\phi$ were prepared by diluting
samples centrifuged to a random close packed sediment, with volume fraction $
\phi_{sed}=\phi_{rcp}$. One traditional method to determine $\phi$ for HS
suspensions employs the crystal-fluid coexistence boundaries [\cite{Segre1995,Pusey1986}].
Because of polydispersity our suspensions do not crystallize, and $\phi$ has
to be determined differently [\cite{Poon2011}]. We have chosen the following method. We first
measure the mass density of the solvent $\rho_{s}$ (densitometer: Anton
Paar DMA 4500). We then measure the density $\rho_{rcp}$ of the close packed sediment
after centrifugation by taking a small sample, diluting it by a factor $
f=10$ by weight in solvent and measuring the density $\rho^*$ of this
diluted suspension. The density of the sediment is then deduced from $
\rho_{rcp}=\rho^*\rho_{s}/[(1+f)\rho_{s}-f\rho^*]$. Samples at a
given $\phi/\phi_{rcp} < 1$ for the main experiments are then prepared by
adding  a solvent mass $M_s$ to the known weight $M_t$ of the stock
sediment (giving $\phi/\phi_{rcp}=[1+\frac{M_s}{M_t}\frac{\rho_{rcp}}{\rho_s}]^{-1}$) and homogenizing the sample rigorously. The densitometry results are given in table~\ref{tab0}; $\rho_{s}$ slightly increases with increasing particle
size, probably due to a slight increase in RI for larger particles
and thus a slight change in the RI-matching composition of the
solvent mixture. We emphasize that, without knowledge of the colloid mass density $\rho_c$, this method does not provide $\phi_{rcp}$  but only $\phi/\phi_{rcp}$. When mentioning absolute volume fractions, we have
assumed $\phi_{rcp}=0.67$, as found in simulations of spheres with a polydispersity of $\sim 15\%$ [\cite{fixschaert}] (monodisperse
spheres give $\phi_{rcp}=0.64$). The value of $\phi_{rcp}=0.67$ may be used to extract the colloid density via
$\phi_{rcp}=(\rho_{rcp}-\rho_s)/(\rho_c-\rho_s)$. The resulting densities $\rho_c \sim 1.12-1.13~$g/ml are reduced compared to the bulk
PMMA value of $\sim 1.2~$g/ml, but reasonable match what is expected from the above mentioned $\sim 10\%$ swelling of the particles
\footnote{We do not separately take into account the reduction of $\rho_c$ arising from the PHS-layer. The density of swollen colloids is then simply $\rho_{c}=(\rho_{PMMA}s^{-3})+[1-s^{-3}]\rho_s$ with $s \simeq 1.1$ the swelling ratio. As the swelling is not exactly known, we have used bare particle radii, measured in decalin, throughout the paper.}. We note that the effect of polydispersity and compaction rate on the value of $\phi_{rcp}$ is still under study [\cite{hermesdijkstra_epl10}] and not without controversy, see also [\cite{Poon2011}]. Yet, in relative terms, our approach is quite accurate, the main uncertainty being a variation $\delta\phi_{rcp}/\phi_{rcp} \simeq 0.005$ found between
different centrifugation runs via drying and weighting. The value of this uncertainty is
close to the maximum variation in the `jamming' fraction for our
polydispersity, as found in simulations of HSs for different compression rates [\cite{hermesdijkstra_epl10}], but our $\delta\phi_{rcp}$ (in two tests we performed, using a single batch) was obtained under similar centrifugation conditions. Moreover, the sedimentation Peclet number  
${\rm Pe}_s = 4 \pi a^4 \Delta\rho g_c / 3k_B T$ (with $\Delta\rho=\rho_c-\rho_s$ and 
$g_c$ the acceleration), lies in the range ${\rm Pe}_s\sim 0.005-1$ from the smallest to the largest particles. Thus, `jamming' effects (which may cause $\phi_{sed}<\phi_{rcp}$) are strongly limited, except possibly for the $a=652~$nm sample. Most important, for the analysis of the $\phi$ dependence of rheological and slip parameters (Sect.~\ref{Sec_slipparam}), we employ the directly measured {\it reduced} volume fraction $\phi/\phi_{rcp}$, which suffers the least from possible ambiguities [\cite{Poon2011}].

\begin{table}[tbp]
\begin{center}
\begin{tabular}{|c|c|c|c|c|}
\hline
sample name & asm340 & asm247 & asm209 & asm195 \\ \hline
$a$ (nm) & 138 & 150 & 300 & 652 \\ \hline
$\rho_{s}$ (g.ml$^{-1}$) & 0.9153 & 0.915 & 0.92023 & 0.92543 \\ \hline
$\rho_{RCP}$ (g.ml$^{-1}$) & 1.0596 & 1.0594 & 1.07 & 1.0644 \\ \hline
\end{tabular}
\end{center}
\caption{Particle size $a$, density of the index-matched solvent $\rho_{s}$ and density of the random close packed sediment $\rho_{RCP}$ for the different samples.}
\label{tab0}
\end{table}

\subsection{Measurement system}

\label{subsec_instr}

\begin{figure}[tbp]
\center\scalebox{1}{\includegraphics{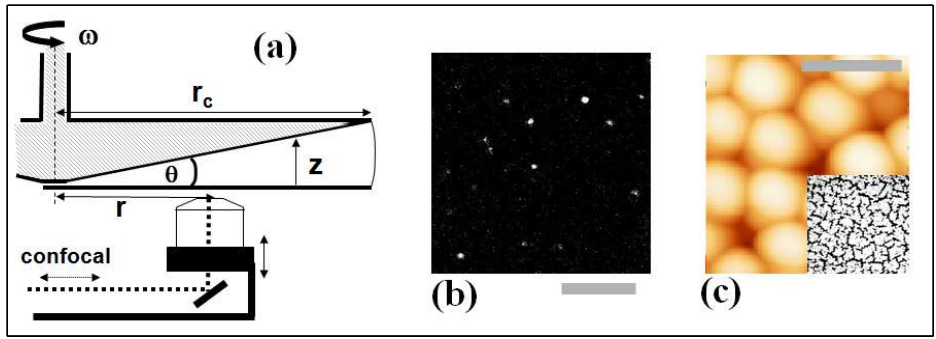}}
\caption{(a) Cone-plate rheometer with transparent plate and optics
connected via an adjustable arm to the confocal scanner. The cone angle $
\protect\theta$ and radius $r_c$ are shown. (b) Confocal image of an
RI-matched suspension with the $a=652$~nm fluorescent tracers showing as
bright spots. Scale bar: $10~\protect\mu$m. (c) AFM image of the sintered layer of $a=652$~nm colloids on the glass slide. The color scale marks height variations of $\sim 500~$nm. Scale bar: $2.5~\protect\mu$m. Inset: confocal image of a coated slide on a larger scale, $50\times50~\protect\mu$m.}
\label{fig_setup}
\end{figure}

We use a stress-controlled rheometer (AR2000, TA Instruments) with a
cone-plate geometry (radius $r_c=20$~mm, cone angle $\theta=1^{\circ}$
unless mentioned otherwise) and a glass slide (radius $25$~mm, thickness $\sim 180$~$\mu$m) as the bottom plate. A solvent trap saturates the
atmosphere around the sample minimizing evaporation. We image the flow with
a confocal scanner (VT-Eye, Visitech International) through the glass plate
via a piezo-mounted objective (oil immersion, magnification $\geq 60\times$
), and optics mounted on an adjustable arm [\cite{Besseling2009}]. We take
2D movies in the flow-vorticity ($x$--$y$) plane at a frame rate $\leq 90$
~Hz at various distances $r$ from the cone center and at equally spaced depths $z \leq h$, with $z$ the velocity-gradient direction and $h$ the local gapsize (Fig.~\ref{fig_setup}(a)), from which we extract,
via image analysis [\cite{Besseling2009}], the velocity profiles $v(z)$ at
different $r$. The typical error bar in the reduced velocity $v(z)/v(h)$ is $\lesssim 5\%$, achieved by imaging sufficiently large displacements, i.e. over timescales of at least a few times $1/\dot\gamma_a$ (see also [\cite{Besseling2010}]). The observation window is 
$\sim 50\times 50$~$\mu$m$^2$, similar to the gap sizes where imaging is performed  ($r=2$--$10$~mm, $h=35$--$175$~$\mu$m). Over the
observation window, the variation in $h$ ($ \delta h \simeq 0.9$~$\mu$m) is
negligible, $\delta h /h \ll 1$, and the geometry locally mimics  
parallel plates separated by $h=r\tan (\theta)$. Experiments
were performed at controlled applied shear rate $\dot\gamma_a$ (using the rheometer's fast feedback), going from
high to low rates, unless stated otherwise. Stress controlled measurements
gave the same results; in particular, we do not find differences between `static' (measured for increasing stress) and dynamic stress thresholds for flow or slip (the latter measured on reducing $\sigma$ or $\dot\gamma_a$), except for possible initial hysteresis just after loading of very concentrated samples (Sect.~\ref{Sec_slipparam}).  

\subsection{Wall properties}

\label{subsec_wallprops}

We have used the following preparations of the surfaces of the cone-plate
geometry: (\textit{i}) To prevent slip, the cone, the glass slide or both
can be made rough on a scale similar to or larger than the particle radius by
spin-coating a $\phi \sim 0.3$ suspension of $a=652~$nm radius particles and
sintering the resulting dense disordered colloidal monolayer for one hour at 
$\sim 120^{\circ}~$C in a vacuum oven. This sintering causes adhesion of the colloids to the glass (or metal cone), probably accompanied by local redistribution of the PHS stabilizer at these temperatures. Nevertheless, the sintering leaves the corrugated nature of the disordered monolayer intact, giving a wall roughness $\sim 500~$nm, see Fig.~\ref{fig_setup}(c) and the inset. The suppression of slip associated with this rough coating has been evidenced in [\cite{Besseling2009}] (Fig. 16 in that paper) and is also shown in Sect.~\ref{sec_localization}. (\textit{ii}) Use of the glass slides
without coating gives a surface which is very smooth on the colloidal scale
(local roughness $<1~$nm measured by AFM ). The slides were used either
untreated or cleaned with ethanol, methanol or a Piranha solution ($98\%$ H$
_2$SO$_4$:$30\%$H$_2$O$_2$ aqueous solution, ratio $7:3$ by mass ), but these
cleaning methods did not give systematic differences in slip behavior. We
also performed temperature-controlled rheological measurements (without
imaging) using the smooth glass slides as bottom plate, achieved by
anchoring the slides on the rheometer Peltier plate using thermal paste. The
stainless steel surface of the cone (cleaned with acetone) is also smooth on
the colloidal scale, but interacts differently with the colloids (Sect.~\ref{sec_wallinteraction}).

\section{Main experimental results}

\label{sec_results}

Rheological studies of concentrated HS suspensions have shown the
emergence of a finite dynamic yield stress for $\phi \gtrsim 0.57-0.58$ [\cite{Petekidis2004, Pham2006}]. For $\phi<\phi_g$, the flow curve, i.e. a plot
of the measured stress ($\sigma_m$) versus applied shear rate ($\dot\gamma_a$%
), exhibits Newtonian behavior at low $\dot\gamma_a$ followed by strong
shear thinning at higher $\dot\gamma_a$, while for $\phi>\phi_g$ suspensions are
glassy and have a yield stress $\sigma_y$ below which no flow occurs. In this case the flow curves exhibit a Herschel-Bulkley behavior
described by $\sigma_m=\sigma_y+\alpha\dot\gamma_a^n$. This behavior is
geometry independent [\cite{Pham2008}], at least for $\phi \lesssim 0.6$,
indicating that $\dot\gamma_a$ and the bulk flow rate of the material, $%
\dot\gamma$, are the same. For larger $\phi$, different geometries may
cause small differences for $\sigma \simeq\sigma_y$ due to shear-banding [%
\cite{Besseling2010}], but the bulk $\sigma(\dot\gamma)$ behavior is
nevertheless consistent with a HB form.

\begin{figure}
\center\scalebox{1}{\includegraphics{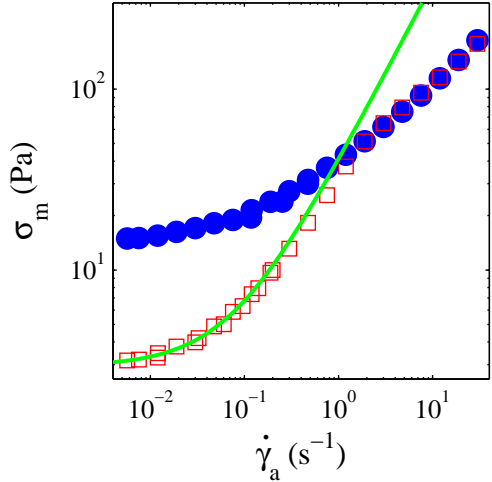}}
\caption{Measured shear stress $\protect\sigma_m$ versus applied shear rate $%
\dot\protect\gamma_a$ for $a=300$~nm colloids at $\protect\phi \sim 0.64$ in
RI-matching solvent ($\square$) and decalin ($\bullet$), using a smooth
glass slide and coated cone. Full line: fit of the low $\dot\protect\gamma_a$
branch to Eq.~(\protect\ref{eq:binghameta}), giving $\protect\eta_{\rm eff}=38.3~ $Pa$\cdot$s, $\protect\sigma_s=2.94~$Pa. The data for decalin are
multiplied by a factor $1.5$ for comparison, the difference with the
RI-matching solvent is due to slightly different $\protect\phi$.}
\label{fig_im_nim}
\end{figure}

We first illustrate how the rheology for $\phi > \phi_g$ is affected when
the colloid-wall interaction is changed. Figure~\ref{fig_im_nim} shows the
stress measured with the rheometer, $\sigma_m$, versus the \textit{applied}
shear rate $\dot\gamma_a$ for a suspension in pure decalin as well as in the
RI-matching decalin-tetralin mixture, at similar volume fractions. For the
RI-mismatched sample, the flow curve clearly shows a yield stress and an
overall Herschel-Bulkley response. However, in the RI-matched sample, the
flow curve at low $\dot\gamma_a$ exhibits a branch with Bingham-like
behavior: 
\begin{eqnarray}
\sigma_m=\sigma_s+\eta_{\mathrm{eff}} \dot\gamma_a.  \label{eq:binghameta}
\end{eqnarray}
Here $\sigma_s$ is a threshold stress below which the (apparent) flow
completely stops and $\eta_{\mathrm{eff}}$ is an effective viscosity
characterizing the stress increase in this branch. For large $\dot\gamma_a$
the stress attains the same nonlinear behavior as the sample in decalin. As shown below,
the small $\dot\gamma_a$ behavior marks full slip along the glass slide (due to the
strongly suppressed colloid-wall vdW attraction) with a vanishing shear rate $\dot\gamma$ in the bulk of the sample.

\begin{figure}
\scalebox{1}{\includegraphics{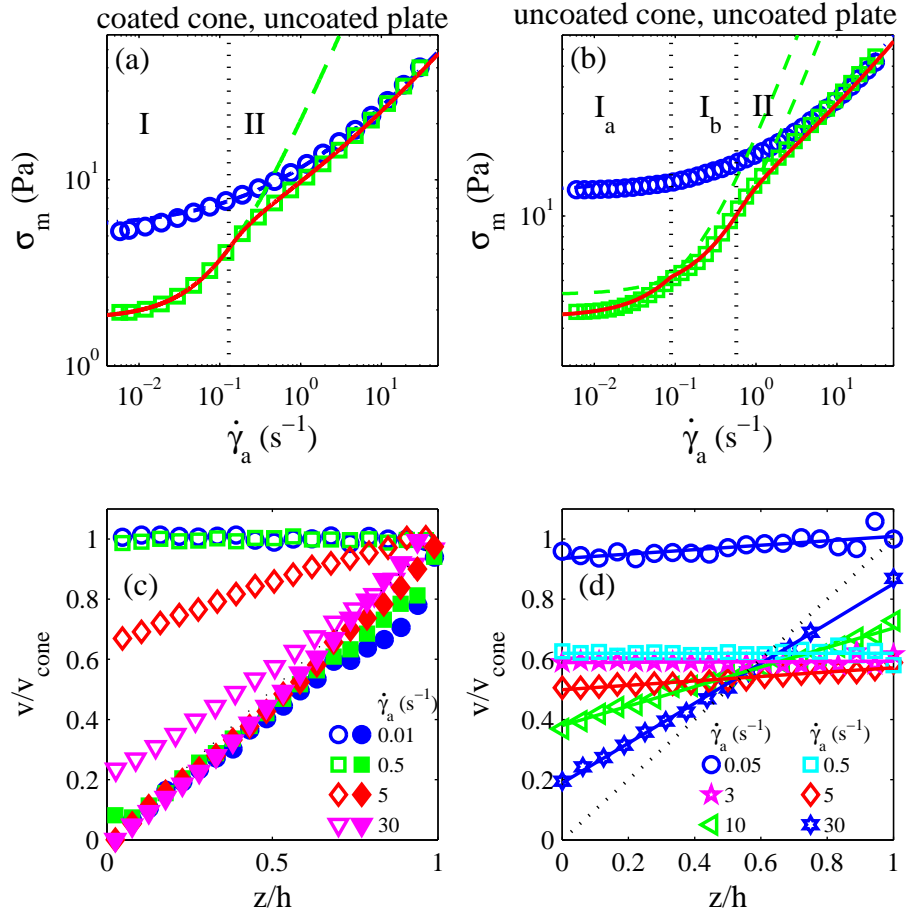}}
\caption{(a,b) Measured stress versus $\dot\protect\gamma_a$ for coated cone
and plate ($\circ$) and with (a) uncoated plate ($\square$) and $\protect\phi%
=0.59$ and $a=138$~nm, and (b) uncoated cone and uncoated plate ($\square$)
with $\protect\phi=0.59$ and $a=150$~nm. Regime $\mathrm{I}$ in (a) and $%
\mathrm{I_a}$ in (b) represent full slip along one boundary; regime $\mathrm{%
I_b}$ in (b) represents full slip along two boundaries; and regime $\mathrm{II}$ in
(a),(b) mark slip plus bulk flow. In (a),(b) the dash-dot curves are
Hershel-Bulkley fits with $n=0.5$, Eq.~(\protect\ref{eq_HB}), giving $%
\protect\alpha=6.1~$Pa$\cdot$s$^{1/2}$, $\protect\sigma_y=5.5~$Pa in (a) and $%
\protect\alpha=8~$Pa$\cdot$s$^{1/2}$, $\protect\sigma_y=14~$Pa in (b); dashed
curves are fits to the Bingham form Eq.~\protect\ref{eq:binghameta}. In
regime $\mathrm{I}$ in (a), where Eq.~(\protect\ref{eq_sigetafI}) applies,
this gives $\protect\beta=8.2 \cdot 10^4~$Pa s/m, $\protect\sigma_s=1.8~$Pa.
In regime $\mathrm{I_a}$ or $\mathrm{I_b}$ in (b), where Eq.~(\protect\ref%
{eq_sigetafI}) and Eq.~(\protect\ref{eq_sigetafIb}) apply, the parameters
are $\protect\beta=8.7.10^4~$Pa$\cdot$s$\cdot$m$^{-1}$, $\protect\sigma_1=5.2~$Pa, $\protect%
\sigma_2=3.4~$Pa; full lines in regime II are the global flow curves from
Eq.~\protect\ref{eq_sigmamtot} in Appendix~\protect\ref{appsec_slipone} using
the above parameters. (c) Normalized velocity profiles $v(z)/v_{cone}$ for
the suspension in (a) at $r=3$~mm with coated surfaces (filled symbols) and
at $r=2.5$~mm with uncoated plate (open symbols) for various $\dot\protect%
\gamma_a$. (d) $v(z)/v_{cone}$ for the data in (b) for uncoated cone and plate, at $r=5.5$~mm and
various $\dot\protect\gamma_a$. Full lines: linear fits. Dotted lines:
behavior without slip.}
\label{fig_globalrheo}
\end{figure}

We next discuss results for different combinations of smooth and colloid-coated surfaces. Figure~\ref{fig_globalrheo} shows flow curves
for RI-matched samples with both walls or one wall coated, Fig.~\ref%
{fig_globalrheo}(a), and with two uncoated walls (smooth glass plate and
stainless steel cone), Fig.~\ref{fig_globalrheo}(b). With both surfaces
coated (Fig.~\ref{fig_globalrheo}(a), open circles), a Herschel-Bulkley
behavior is recovered, confirming that coating prevents slip, as we also
directly observed via microscopy [\cite{Besseling2007,Ballesta2008}]. If only
one surface is coated (squares in Fig.~\ref{fig_globalrheo}(a)), we observe
a similar slip response as described above: below a critical applied shear
rate $\dot\gamma_{a,c}$ bulk shear vanishes in the entire gap (see
below and the appendix) and we have a Bingham response; we define this as
regime I. For $\dot\gamma_a>\dot\gamma_{a,c}$ the flow curve deviates from
this Bingham regime and approaches the flow curve obtained with coated geometry, this regime is noted regime
II. If both walls are uncoated, squares in Fig.~\ref{fig_globalrheo}(b), the
flow curve exhibits two successive Bingham regimes at low shear rate: $%
\mathrm{I_a}$ marking slip at the plate, $\mathrm{I_b}$ marking slip at both
surfaces (see Sec.~\ref{Sec_model}); the second Bingham regime, $\mathrm{I_b%
}$, has a higher slip stress and an effective viscosity $\eta_{\mathrm{eff}}$
half the value of that in regime $\mathrm{I_a}$. Eventually, at high shear
rate, the curve again tends toward the HB flow curve.

We now turn to the velocimetry results obtained by simultaneous confocal
imaging of the flow, Fig.~\ref{fig_globalrheo}(c,d). In Fig.~\ref%
{fig_globalrheo}(c) the filled symbols show flow profiles with both walls
coated. These profiles are nearly linear and $v(z)$ reaches zero at the
glass plate and the applied velocity at the cone, showing that the coating
provides a no-slip condition. The open symbols in Fig.~\ref{fig_globalrheo}%
(c),(d) show $v(z)$ for one uncoated surface (c) or two uncoated surfaces
(d). In both cases $v(z)$ is essentially linear but exhibits slip at the
non-coated walls. The profiles can be fitted by $v=v_s+\dot\gamma z$, with $%
v_s$ the slip velocity at the plate and $\dot\gamma$ the bulk shear rate. At
the smallest $\dot\gamma_a$, $\dot\gamma$ is zero; the suspension sticks to
the coated cone and rotates as a solid body, slipping over the smooth glass.
This causes the apparent flow below the yield stress in regime $\mathrm{I}$
in Fig.~\ref{fig_globalrheo}(a) or $\mathrm{I_a}$ in (b). Figure~\ref%
{fig_globalrheo}(c) shows that plug flow can persist in regime II for $%
r=2.5~ $mm and $\dot\gamma_a=0.5~$s$^{-1}$, as explained later. With two
uncoated surfaces, Fig.~\ref{fig_globalrheo}(d), the velocity profiles for
intermediate $\dot\gamma_a$ show slip at both walls, with solid body
rotation at a fraction of the cone velocity. This occurs in regime $\mathrm{%
I_b}$ and, as shown later, also in regime $\mathrm{II}$
at small $r$. Eventually, for applied rates $\dot\gamma_a \geq 5~$s$^{-1}$,
both in Fig.~\ref{fig_globalrheo}(c) and (d) the bulk of the sample starts
to yield, $\dot\gamma>0$. The difference between $\dot \gamma$ and $%
\dot\gamma_a$ decreases on increasing $\dot\gamma_a$ and the flow curve
approaches bulk HB behavior.

\begin{figure}
\scalebox{1}{\includegraphics{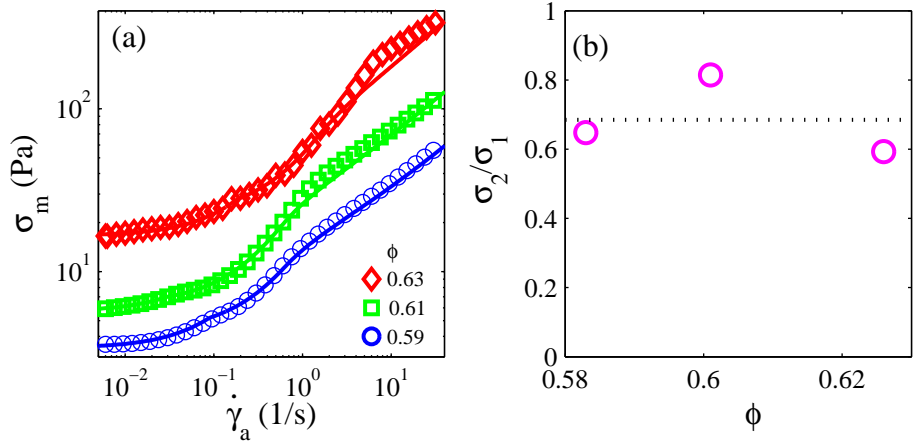}}
\caption{(a) Flow curves for two uncoated surfaces, $a=150$~nm, and various $%
\protect\phi$. The lower branches of the full lines are fits to Eqs.~(%
\protect\ref{eq_sigetafI},\protect\ref{eq_sigetafIb}), the upper branch is
Eq.~\protect\ref{eq_sigmamtot} for slip at two surfaces. The parameters are $%
\protect\sigma_{1,2}=[7,5.7]~$Pa, $\protect\beta=19.0 \cdot 10^4~$Pa$\cdot$s$\cdot$m$^{-1}$, $%
\protect\sigma_y=23.5~$Pa, $\protect\alpha=18.0~$Pa$\cdot$s$^{1/2}$ for $\protect%
\phi=0.61$ and $\protect\sigma_{1,2}=[28,16.6]~$Pa, $\protect\beta=26.3\cdot
10^4~$Pa$\cdot$s$\cdot$m$^{-1}$, $\protect\sigma_y=82~$Pa, $\protect\alpha=50.6~$Pa$\cdot$s$^{1/2}$
for $\protect\phi=0.63$ (the $\phi=0.59$ parameters are as in Fig.~\ref{fig_globalrheo}b). (b) Ratio of the slip threshold stresses at the
bottom and top surface, $\protect\sigma_2/\protect\sigma_1$, versus $\protect%
\phi$, the dotted line is the average value.}
\label{fig_op2slip}
\end{figure}

\begin{figure}
\center\scalebox{1}{\includegraphics{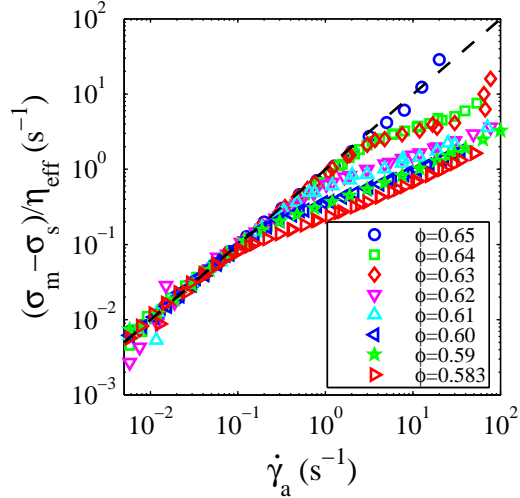}}
\caption{Shear stress minus slip stress $\protect\sigma_s(\protect\phi)$
divided by the Bingham viscosity $\protect\eta_{\mathrm{eff}}(\protect\phi)$
for $a=138$~nm and $\protect\phi=0.585$--$0.65$. $\protect\sigma_s$ and $%
\protect\eta_{\mathrm{eff}}$ were extracted from fits of the small $\dot%
\protect\gamma_a$ behavior to Eq.~(\protect\ref{eq:binghameta}). Dashed
line: $\protect\sigma_m-\protect\sigma_s=\dot\protect\gamma_a \protect\eta_{%
\mathrm{eff}}$.}
\label{powerlawone}
\end{figure}

The typical behavior described above, with the transition from a Bingham to
a HB flow regime, is representative for all RI-matched suspensions with $%
\phi > 0.57$ and at least one smooth wall. We illustrate this in Fig.~\ref%
{fig_op2slip} and Fig.~\ref{powerlawone}. The former shows the two-step
Bingham behavior for two uncoated walls, as in Fig.~\ref{fig_globalrheo}(b)
but now including two other concentrations. Both the slip threshold stress ($%
\sigma_1 $ for the cone, $\sigma_2$ for the plate) and $\sigma_y$ increase
strongly with $\phi$. Figure.~\ref{fig_op2slip}(b) shows that the ratio
between the slip threshold stress for the two surfaces is essentially
independent of $\phi$. In Fig.~\ref{powerlawone} we show $%
(\sigma-\sigma_s)/\eta_{\mathrm{eff}}$ versus applied shear rate for
suspensions with $a=138$~nm at various $\phi$, using a coated cone and
smooth glass plate. Plotted this way, all curves at small $\dot\gamma_a$
(regime I) are linear and overlap, following Eq.~(\ref{eq:binghameta}). In
addition to the threshold stress, also the fitted values of the Bingham
viscosity show a strong $\phi$-dependence, both of which are discussed in Sec.~\ref%
{Sec_slipparam}.

\begin{figure}
\scalebox{1}{\includegraphics{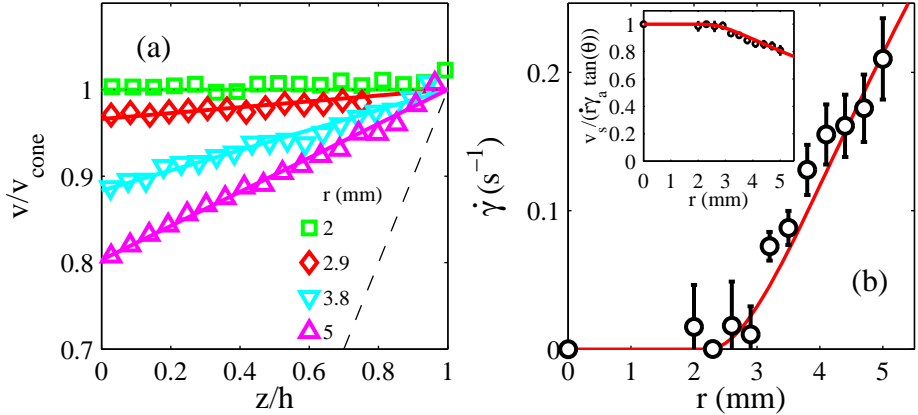}}
\caption{(a) Velocity profiles for fixed shear rate $\dot\protect\gamma%
_a=1.1 $~s$^{-1}$, $\protect\phi=0.59$ and $a=138$~nm with uncoated glass
and coated cone, at selected distances $r$ from the center. (b)
Corresponding local shear rate versus $r$. Full line: Eq.~(\protect\ref%
{eq_1u_dg_dga}) with $h=r \tan(\protect\theta)$, using Eq.~(\protect\ref%
{eq_param1slip}) and rheological parameters as in Fig.~\protect\ref%
{fig_globalrheo}(a). Inset: corresponding $r$-dependence of the normalized slip velocity $\tilde{v}_s=v_2/(\dot\gamma r \tan{\theta})$, along
with the prediction using Eqs.~(\protect\ref{eq_ratevslip}),(\protect\ref{eq_1u_dg_dga}) and (\protect\ref{eq_param1slip}) with $v_1=0$
(full line).}
\label{fig_localrdep}
\end{figure}

Figure~\ref{fig_localrdep}(a) shows velocity profiles from the experiment in
Fig.~\ref{fig_globalrheo}(a) ($\phi=0.59$, coated cone, uncoated plate), but
at a constant applied shear rate $\dot\gamma_a=1.1$~s$^{-1}$, at various
distances $r$. For this applied rate, $\sigma_m$ is close to its value
measured in the absence of slip (see Fig.~\ref{fig_globalrheo}(a)), but
slip is nevertheless visible. For large $r$, slip and bulk flow are
simultaneously present, the mechanism governing slip in regime I thus
persists when bulk shear is present. The slip gets more pronounced for
smaller $r$, which is shown directly in Fig.~\ref{fig_localrdep}(b) via the $%
r$-dependence of the local shear rate $\dot\gamma(r)$ extracted from Fig.~%
\ref{fig_localrdep}(a): $\dot\gamma(r)$ decreases strongly with $r$ and
eventually vanishes completely for $r< r_y \simeq 2.5~$mm, with $r_y$ a
``yielding radius'' inside of which no shear is present. These data demonstrate
directly that, in presence of slip, the stress in a cone-plate, while
essentially uniform across the gap, is radially non-uniform: for small $r$, $%
\sigma(r) < \sigma_y$, while for larger $r$, $\sigma(r)>\sigma_y$.

The velocity profiles at small $\dot\gamma_a$ in Figs.~\ref{fig_globalrheo} suggest that
in regime $\mathrm{I}$ the suspension moves as a solid body, without bulk
shear, down to the glass plate. To get direct evidence for this and exclude
locally non-uniform flow as seen in colloidal crystals [\cite%
{Derks2004,Cohen2006}], we imaged the first layers of a suspension of fully
fluorescent particles ($a=652$~nm) in regime I in the RI-matching solvent.
As reported in Fig. 2a of [\cite{Ballesta2008}], in regime $\mathrm{I}$
below the critical applied shear rate $\dot\gamma_{a,c}$ mentioned earlier,
we observed plug flow with $v_s=v_{cone}$ down to the first layer of
particles. There is thus no highly sheared layer of colloids and the Bingham
slip response, Eq.~\ref{eq:binghameta}, reflects the behavior of a
lubrication layer between the first particles and the wall. Due to the
smooth wall, the colloids do exhibit some layering at the surface, but this
vanishes completely beyond $\sim 4$ particle diameters (Fig. 2a of [\cite{Ballesta2008}%
]), where the structure is that of a fully disordered colloidal glass. Even near the wall, no significant ordering was observed within the imaging plane.

\section{Model for the slip and yield behavior}

\label{Sec_model}

The local and global rheology in the previous section can be described via a
model based on the same assumption used in [\cite{Yoshimura1988, Russel2000,
Kalyon2005}], i.e. that the stress at the wall due to slip matches the
stress in the bulk. As suggested by the phenomenology of the smooth wall
flow curves, Fig.~\ref{fig_globalrheo}(a),(b) and the Bingham form Eq.~\ref%
{eq:binghameta} for the slip branch, the local stress $\sigma_{\mathrm{slip}%
} $ and the slip velocity $v_i$ of the colloids along a wall, labeled by $i$%
, can be related via: 
\begin{eqnarray}
\sigma_{\mathrm{slip}}=\sigma_i + \beta_i v_i.  \label{eq_slipstress}
\end{eqnarray}
Here $\beta_i v_i$ is a hydrodynamic term reflecting the lubrication between
the first layer of colloids and the wall and the threshold stress $\sigma_i$
is similar to a Coulomb friction term. Unless particles strongly interact
with one of the smooth surfaces the mean distance of a particle to the
surface will mostly be determined by the available free space, which only
depends on $\phi$; hence we assume that $\beta_1=\beta_2=\beta(\phi)$. On
the other hand we expect $\sigma_i$ to depend on the wall interaction which
can differ between the surfaces. A more detailed description of these terms
will be given later. Below we use the convention $i=1$ for the top
surface, $i=2$ for the bottom plate and also assume that $\sigma_1\geq \sigma_2$.

For the bulk flow, we use the HB form, taking for simplicity an exponent $n=0.5$: 
\begin{eqnarray}
\sigma_{\mathrm{bulk}}=\sigma_y + \alpha \dot\gamma^{0.5}.  \label{eq_HB}
\end{eqnarray}
Although the measured $n$ can vary ($0.35 \lesssim n \lesssim 0.6$) for $1
\gtrsim \phi/\phi_{rcp} \gtrsim 0.85$, as long as $n<1$ the results below
remain qualitatively correct. The relative contribution of slip and flow is
then evaluated via $\sigma_{\mathrm{bulk}}=\sigma_{\mathrm{slip}}$ along
with the relation between $\dot\gamma$ and $v_{1,2}$: 
\begin{eqnarray}
h\dot\gamma+v_1+v_2=h \dot\gamma_a.  \label{eq_ratevslip}
\end{eqnarray}
Due to the dependence on the gap size $h$, the results are geometry
dependent. We illustrate this for infinite parallel plates and the
cone-plate geometry.

\subsection{Infinite parallel plates}

This idealized situation serves as the basis for subsequent
analysis of the cone-plate geometry. Various cases are possible depending on
the relative values of $\sigma_s$, $\sigma$ and $\sigma_y$. We first
consider the case $\sigma <\sigma_{y}$. In this regime bulk shear is absent, 
$\dot\gamma=0$. For $\sigma_2\leq \sigma <\sigma_1, \sigma_y$ (regime $\mathrm{I}$
or $\mathrm{I_a}$) slip is localized at one surface, leading to $%
\sigma=\sigma_2 +\beta h\dot\gamma_a$. For $\sigma_1,\sigma_2 \leq
\sigma<\sigma_{y}$ (regime $\mathrm{I_b}$) the sample slips at both surfaces
and we have $\sigma = (\sigma_1+\sigma_2)/2+\beta h\dot\gamma_a/2$.

Whenever $\sigma_2<\sigma_y$, yielding sets in at a slip velocity $%
v_2=(\sigma_y-\sigma_2)/\beta$. This defines the transition from regime I or 
$\mathrm{I_b}$ to regime II, where bulk flow and slip coexists. The relation 
$\dot\gamma(\dot\gamma_a)$ is obtained by equating Eqs.~(\ref{eq_slipstress}) and ({\ref{eq_HB}), using Eq.~(\ref{eq_ratevslip}). For slip at one wall ($%
v_1=0 $, $\sigma_2\leq\sigma_y\leq \sigma <\sigma_1$), this gives: 
\begin{eqnarray}  \label{eq_1u_dg_dga}
\dot\gamma =\dot\gamma_a -\dot\gamma_y +\dot\gamma_0\left(1- \sqrt{1+\frac{2%
}{\dot\gamma_0}\left(\dot\gamma_a -\dot\gamma_y \right)}\right),
\end{eqnarray}
\begin{eqnarray}  \label{eq_param1slip}
\Delta\sigma =\sigma_y -\sigma_2 \mbox{, } \dot\gamma_y=\frac{\Delta\sigma}{%
\beta h} \mbox{, and } \dot\gamma_0=\frac{\alpha^2}{2h^2\beta^2}.
\end{eqnarray}
For $\sigma_{1,2}<\sigma_y<\sigma$, the sample yields and slips at both
surfaces. The relation $\dot\gamma(\dot\gamma_a)$ is still given by Eq.~(\ref%
{eq_1u_dg_dga}), with: 
\begin{eqnarray}  \label{eq_param2slip}
\Delta\sigma =\sigma_y -\frac{\sigma_1 +\sigma_2}{2} \mbox{, } \dot\gamma_y=%
\frac{2\Delta\sigma}{\beta h} \mbox{, and } \dot\gamma_0=\frac{2\alpha^2}{%
h^2\beta^2}.
\end{eqnarray}

Equation~(\ref{eq_1u_dg_dga}) can be written in dimensionless form, giving
the following master curve for the local shear rate versus applied rate in
regime II: 
\begin{eqnarray}
\dot\Gamma=1+\Omega-\sqrt{1+2\Omega},  \label{eq_mastercurve}
\end{eqnarray}
with $\dot\Gamma=\dot\gamma/\dot\gamma_0$ and $\Omega=(\dot\gamma_a-\dot%
\gamma_y)/\dot\gamma_0$, in which the $h$ dependence is absorbed. This
form describes the vanishing bulk shear rate for $\dot\gamma_a \rightarrow
\dot\gamma_y$ and the approach towards the HB curve ($%
\dot\gamma\rightarrow \dot\gamma_a$) for large rate. This can also be
described in terms of the slip length $l_s$, see App.~\ref{subannexe_ls};
entering regime II, $l_s$ decreases and approaches $l_s \sim
\dot\gamma^{-1/2}$ for large $\dot\gamma_a$.

\subsection{Cone and plate}

\label{sec_coneplate}

To analyze the cone-plate geometry, the variation in gap size needs to be
accounted for. Approximating the geometry at a distance $r$ by parallel
plates, the local stress is deduced as before, with $%
\sigma(r)=\sigma(h/\tan(\theta),\dot\gamma_a)$. The measured stress $%
\sigma_m $ is found by integrating $\sigma(r)$ over the entire geometry: 
\begin{equation}
\sigma_m=\frac{1}{\pi r_c^2}\int_0^{r_c}\sigma(r)2\pi rdr.
\label{eq_stressint}
\end{equation}
We again first consider the case $\sigma(r) <\sigma_{y}$, i.e. solid body
rotation of the entire sample ($\dot\gamma=0$ for all $r$). In regime $%
\mathrm{I}$ or $\mathrm{I_a}$, where slip occurs at the plate ($\sigma_2\leq
\sigma <\sigma_1,\sigma_y$) we have: 
\begin{equation}
\sigma_m^{\mathrm{I}}=\sigma_2+2\beta \tan(\theta) r_c \dot\gamma_a /3.
\label{eq_sigetafI}
\end{equation}
which is the Bingham form Eq.~({\ref{eq:binghameta}) with $\eta_{\mathrm{eff}%
}=2\beta \tan(\theta) r_c/3$. For complete slip at two surfaces $%
\sigma_1,\sigma_2\leq \sigma <\sigma_{y}$, regime $\mathrm{I_b}$, the solid body
rotation rate of the sample is reduced, see Eq.~(\ref{eq_omega_Ib}) in App.~%
\ref{appsec_sliptwo}. The stress is obtained from a balance of the total
stress on the bottom plate and the cone (App.~\ref{appsec_sliptwo}): 
\begin{equation}
\sigma_m^{\mathrm{I_b}}=(\sigma_1+\sigma_2)/2+\beta \tan(\theta) r_c
\dot\gamma_a /3.  \label{eq_sigetafIb}
\end{equation}
The transition I$\mathrm{_a} \rightarrow$ I$\mathrm{_b}$ occurs when $%
\sigma_m^{\mathrm{I_a}}=\sigma_m^{\mathrm{I_b}}$ giving Eq.~(\ref%
{eq_omega_Iab}) in App.~\ref{appsec_sliptwo}. }

\begin{figure}
\center\scalebox{1}{\includegraphics{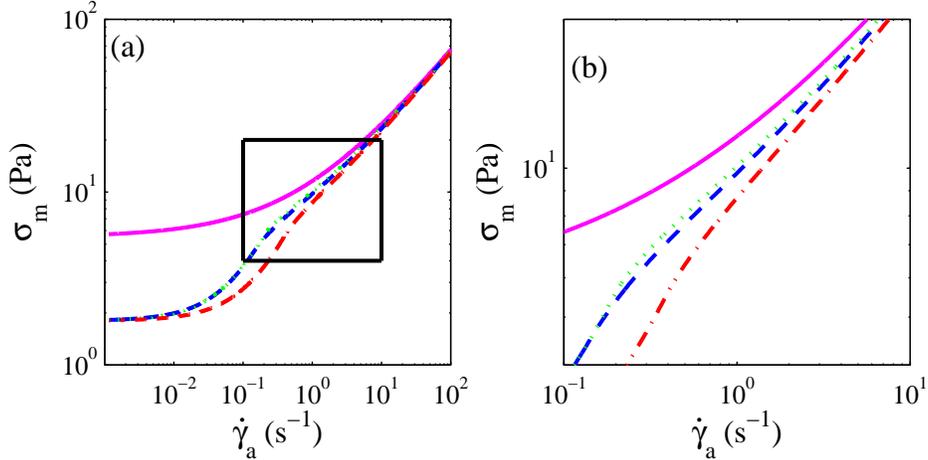}}
\caption{(a) Full line: HB flow curve with $n=0.5$, $\protect\alpha=6.1$~Pa.s
$^{-0.5}$, and $\protect\sigma_y=5.5$~Pa. Other curves: calculations for
different geometries with slip at the bottom plate using $\protect\beta%
=82000 $~Pa.s$/$m, $\protect\sigma_s=1.8$~Pa: cone-plate with $\protect\theta%
=1^{\circ}$, $r_c=20$~mm (dashed line, Eqs.~(\protect\ref{eq_sigetafI},%
\protect\ref{eq_sigmamtot}); parallel plate with $h=2\tan(\protect\theta%
)r_c/3$ (dotted) and $h=\tan(\protect\theta)r_c/3$ (dash-dot),
using Eqs.~(\protect\ref{eq_HB},\protect\ref{eq_1u_dg_dga},\protect\ref%
{eq_param1slip}). (b) Zoom in on (a).}
\label{fig_compare}
\end{figure}

Due to the non-uniform slip velocity and  associated nonuniform
stress, the yielding transition and flow in regime II differ from that
for parallel plates. For slip at the bottom plate ($\sigma_1=\infty$, $v_1=0$%
), when increasing $\dot\gamma_a$, yielding starts when $\sigma_{%
\mathrm{slip}}=\sigma_y$, i.e. at a radius $r_y=(\sigma_y-\sigma_2)/(\beta%
\tan(\theta) \dot\gamma_a)$. The applied rate where regime II starts follows
from $r_y=r_c$: 
\begin{equation}  \label{eq_dotgamac}
\dot\gamma_{a,c}= (\sigma_y-\sigma_2)/\beta r_c \tan(\theta)
\end{equation}
For $\dot\gamma_a>\dot\gamma_{a,c}$, the boundary
between solid body rotation ($r<r_y$) and slip and shear ($r>r_y$) moves
inward. The measured stress in regime II follows from Eq.~\ref{eq_stressint}
integrated over these two regions. The result, Eq.~(\ref{eq_sigmamtot}) in
App.~\ref{appsec_slipone} is shown for specific parameters in Fig.~\ref%
{fig_compare}, along with that for parallel plates for two different gaps.
The transition I-II for the cone-plate is smoother than for parallel
plates due to the 'mixed' nature of the transition in the former.

For slip at two surfaces, the transition to regime II is somewhat more
complicated due to a small difference in $\sigma(r)$ between bottom plate
and cone in regime I. Yet, with the approximation $\sigma_1-\sigma_2 \ll
\sigma_y$ the analysis is essentially the same as for slip at one surface,
see App.~\ref{appsec_sliptwo}: the transition to regime II is described by
Eq.~(\ref{eq_dotgamac}) with the substitution $\sigma_2 \rightarrow
(\sigma_1+\sigma_2)/2$ and $\beta \rightarrow \beta/2$.

\section{Comparison with experiment}

\label{sec_compare}

\subsection{Global rheology}

\label{checkglobal}

The uniform lubrication layer and solid body rotation in regime I lead to an
effective viscosity $\eta_{\mathrm{eff}} \sim \beta r_c \tan (\theta)$,
which can be verified by comparing slip branches for different cones. Figure~%
\ref{fig_geo}(a) shows results for coated cones with different $r_c$ and $%
\theta$ and uncoated glass plate for the same sample ($a=138$~nm, $\phi=0.63$%
). The data for $r_c=20~$mm, $\theta=2^{\circ}$ and $r_c=10~$mm, $%
\theta=4^{\circ}$ indeed superimpose for all $\dot\gamma_a$, as all geometry
dependence of $\sigma_m$ in regime I and II (Eq.~(\ref{eq_sigmamtot}))
enters via $r_c\tan(\theta)$. When plotted versus the velocity at the edge
of the geometry, $\dot\gamma_a r_c \tan (\theta)$, Fig.~\ref{fig_geo}(b),
all slip branches superimpose.

\begin{figure}
\center\scalebox{1}{\includegraphics{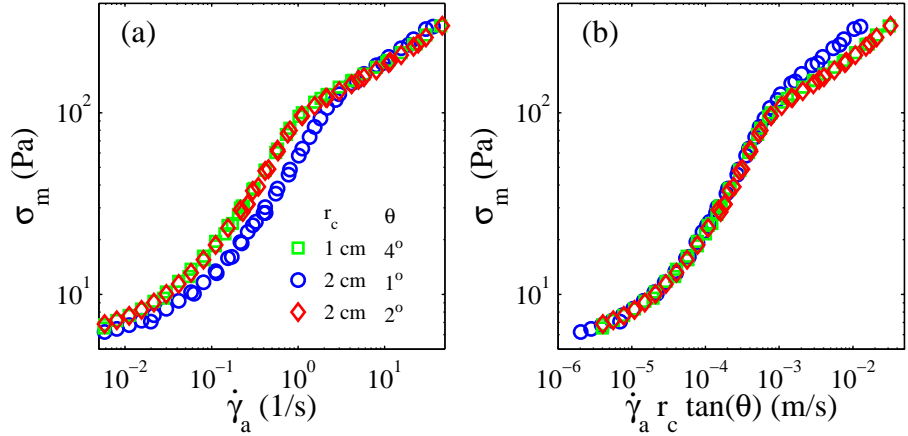}}
\caption{(a) Flow curves for an RI-matched suspension with $a=138$~nm, $%
\protect\phi=0.63$, using different coated cones and smooth plates. (b) $%
\protect\sigma_m$ versus velocity at the geometry edge; symbols as in (a).}
\label{fig_geo}
\end{figure}

The transition from Bingham slip to HB bulk flow behavior in the global flow
curves is well described by Eqs.~(\ref{eq_sigetafI}, \ref{eq_sigetafIb}) and
(\ref{eq_sigmamtot}). The parameters $\alpha$, $\sigma_y$ and $\beta$ and $%
\sigma_{1,2}$ entering Eq.~(\ref{eq_sigmamtot}) (for regime II) follow from
fits of the global rheology with and without coating. Examples of the
predicted full flow curves are shown in Fig.~\ref{fig_globalrheo}(a)(b) and
Fig.~\ref{fig_op2slip}. In Fig.~\ref{fig_globalrheo}(a) the measured
transition from regime I to II for one smooth wall is well captured. In Fig.~%
\ref{fig_globalrheo}(b) and ~\ref{fig_op2slip}, where both walls are
uncoated, the transition from slip at one surface to two surfaces and the
subsequent yielding are also well described. The small discrepancy for $%
\phi=0.63$ can be attributed to both a slight reduction of the HB exponent $%
n<0.5$ for the largest $\phi$ and the presence of shear localization
near the yield stress in this $\phi$ regime, see Sec.~\ref%
{sec_localization}.

\begin{figure}
\scalebox{1}{\includegraphics{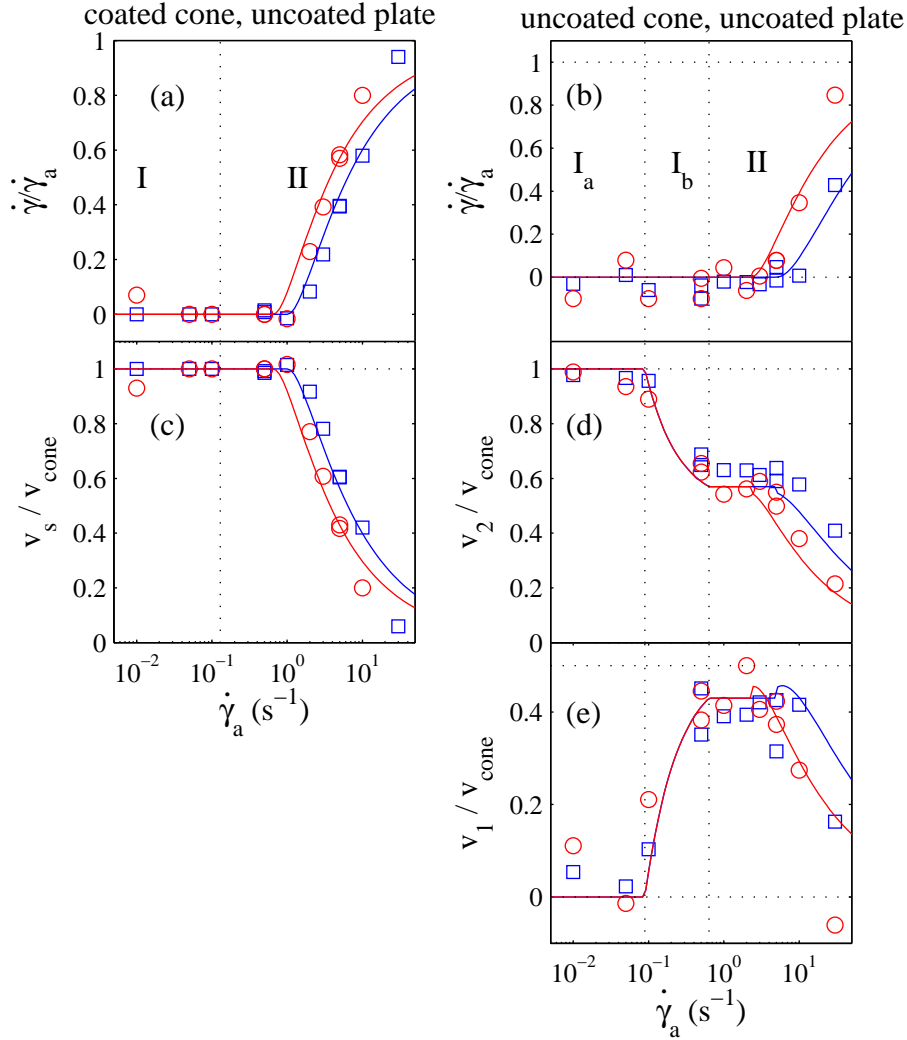}}
\caption{Velocimetry results corresponding to the data in Fig.~\protect\ref%
{fig_globalrheo}. (a),(b) Measured normalized local shear rate versus $\dot%
\protect\gamma_a$ for (a) $r=2.5$~mm ($\square$) and $r=4$~mm ($\circ$), and
(b) $r=2.5$~mm ($\square$) and $r=5.5$~mm ($\circ$). (c),(d) Normalized slip
velocity $v_s/v_{cone}$ at the glass plate corresponding to data in (a),(b).
In (e) the normalized slip velocity at the cone is shown. Data in (a),(c)
are for $a=138$~nm, coated cone and uncoated plate. Data in (b),(d) and (e)
are for $a=150$~nm, uncoated cone and uncoated plate. Full lines in (a),(b)
for $\dot\protect\gamma>0$ are given by Eqs.~(\protect\ref{eq_1u_dg_dga},%
\protect\ref{eq_param1slip}) and Eqs.~(\protect\ref{eq_1u_dg_dga},\protect
\ref{eq_param2slip}), respectively, with parameters given in the
caption of Fig.~\protect\ref{fig_globalrheo}. In (d),(e), the curves in
regime I$\mathrm{_b}$ are given by Eq.~(\protect\ref{eq_omega_Ib}); the
transition I$\mathrm{_a} \rightarrow$ I$\mathrm{_b}$ occurs at $\dot\protect%
\gamma_a^*$ given in Eq.~(\protect\ref{eq_omega_Iab}). The bulk (slip)
velocity at the second $v_{1,2}$ plateau (corresponding to $r<r_y$ in regime
II) is given by Eq.~(\protect\ref{eq_rotvelunshear}); In (c) and (d),(e) the
curves for the largest $\dot\protect\gamma_a$ (where $r>r_y$) follow from
those in (a),(b) via Eq.~(\protect\ref{eq_ratevslip}).}
\label{fig_localrheo}
\end{figure}

\subsection{Local shear rate}

\label{checklocal}

The model can be checked further by comparing flow profiles to
the predictions for local shear rate and slip velocities in Sec.~\ref%
{Sec_model}. To obtain these quantities, $v(z)$ is fitted to 
$v=\dot\gamma z + v_s$. Figure~\ref{fig_localrheo}(a),(c) shows the
reduced bulk rate $\dot\gamma/\dot\gamma_a$ and slip velocity $v_s$ at the
plate from $v(z)$ taken at $r=2.5$~mm (shown in Fig.~\ref%
{fig_globalrheo}(c)) and $r=4~$mm. As expected from Eq.~(\ref{eq_1u_dg_dga}%
), bulk shear starts at smaller $\dot \gamma_a$ for larger $r$, hence for $%
r<r_c$ the solid body rotation extends beyond the transition rate $%
\dot\gamma_{a,c} \simeq 0.135~$s$^{-1}$ given by Eq.~(\ref{eq_dotgamac}).
Similarly Fig.~\ref{fig_localrheo}(b-d-e) presents the local bulk rate and
slip velocities $v_1(\dot\gamma_a)$, $v_2(\dot\gamma_a)$ for uncoated cone
and plate taken at $r=5.5$~mm (from Fig.~\ref{fig_globalrheo}%
(d)) and at $r=2.5~$mm. Both the transition from regime I$\mathrm{_a}$ to I$\mathrm{_b}$  and that to regime II are well described. Due to the small difference between the slip stresses $\sigma_1$ and $\sigma_2$ for bottom
and top, the extent of regime $\mathrm{I_b}$ is limited. Note that $v_1$ and 
$v_2$ are essentially constant between $\dot\gamma_{a,c}$ and $%
(2\sigma_y-\sigma_1-\sigma_2)/\beta r\tan(\theta)\simeq 4-7~$s$^{-1}$, which
motivated the simplification of the model for two uncoated plates in App.~%
\ref{appsec_sliptwo}. For $\dot \gamma_a \gtrsim 4-7~$s$^{-1}$, the bulk
shear rate $\dot\gamma$ and $v_{1,2}$ also match the predictions.

\begin{figure}
\scalebox{1}{\includegraphics{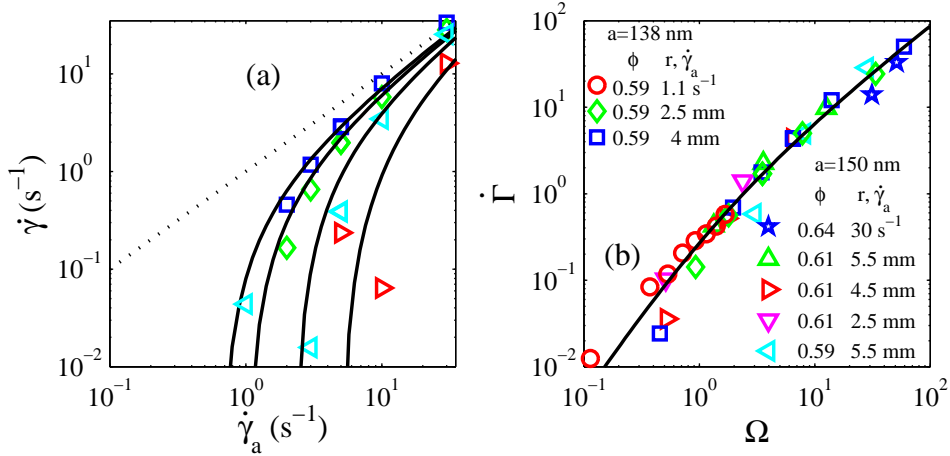}}
\caption{(a) Local shear rate $\dot\protect\gamma$ extracted from the
measured $v(z)$, versus $\dot\protect\gamma_a$ for different $\protect\phi$, 
$a$ and $r$, see symbols in (b), for slip at the bottom plate only ($a=138~$%
nm data) and at both surfaces ($a=150~$nm data). The full lines represent
Eq.~(\protect\ref{eq_1u_dg_dga}) with the rheological parameters entering
via Eq.~(\protect\ref{eq_param1slip}) or Eq.~(\protect\ref{eq_param2slip})
with $h=r\tan(\protect\theta)$. Dotted line: $\dot\protect\gamma%
=\dot\protect\gamma_a$. (b) Normalized local shear rate $\dot\Gamma$ versus
normalized applied rate $\Omega$ for various $r$, $\protect\phi$ and two
particle sizes. The full line shows Eq.~(\protect\ref{eq_mastercurve}).}
\label{fig_mcurve}
\end{figure}

The radial variation of $\dot \gamma$ over the geometry in regime II,
shown in Fig.~\ref{fig_localrdep}(b), is also well explained by the model.
The line in Fig.~\ref{fig_localrdep}(b) shows that
both the onset of yielding at $r_y\simeq (\sigma_y-\sigma_s)/(\beta
\theta\dot\gamma_a)=2.35~$mm and the $r$-dependence of the bulk shear rate
directly follow from the measured rheological parameters. Finally, in Fig.~%
\ref{fig_mcurve}(a) we show $\dot\gamma$ versus $\dot\gamma_a$ for different
$\phi$ and particle sizes along with the model predictions (see
caption). Using the normalization described below Eq.~\ref{eq_mastercurve}, $%
\dot\Gamma=\dot\gamma/\dot\gamma_0$ and $\Omega=(\dot\gamma_a-\dot\gamma_y)/%
\dot\gamma_0$, all data indeed collapse on the master curve $%
\dot\Gamma=1+\Omega-\sqrt{1+2\Omega}$, Fig.~\ref{fig_mcurve}(b). Overall,
both macroscopic and microscopic observations thus validate the
phenomenological model.

\section{Analysis of bulk flow and slip parameters}

Before discussing the nature of the slip behavior, we first present the bulk rheological parameters, measured in detail for the $a=138~$nm and $a=150~$nm samples using rough, coated walls. Figure~\ref{fig_bulkrheovsphi}(a) shows that $\sigma_y$, plotted as function of the normalized distance to random close packing, strongly increases towards rcp and is reasonably described by $\sigma_y a^3/k_B T \simeq 0.01 (1-\phi/\phi_{rcp})^{-3}$. This increase of $\sigma_y$ is due to tightening of particle cages and strongly increasing entropic barriers. The HB exponent $n$, Fig.~\ref{fig_bulkrheovsphi}(b), is in the range $0.4-0.55$ with a slightly decreasing trend for $\phi \rightarrow \phi_{rcp}$ \footnote{These exponents do not necessarily reflect terminal slopes of $\log(\sigma_m)$ vs $\log(\dot\gamma_a)$. Somewhat larger values ($n\lesssim 0.65$) have also been observed in other PMMA HS suspensions [\cite{Koumakis2012}]}. Further, the HB parameter $\alpha$ also strongly increases with $\phi$, Fig.~\ref{fig_bulkrheovsphi}(c), following approximately $\alpha \propto (1-\phi/\phi_{rcp})^{-2.5} \simeq (1-\phi/\phi_{rcp})^{n-p}$. This implies the following scaling of the bulk HB rheology: 
\begin{equation}
\sigma=\sigma_y+\alpha \dot\gamma^n=\sigma_y(\phi)[1+ A(1-\phi/\phi_{rcp})^n (\dot\gamma \tau_B)^n]=\sigma_y(\phi)[1+A(\dot\gamma \tau_m(\phi))^n], \label{eq:HB scaling}
\end{equation}
\begin{figure}[!h]
\scalebox{1.4}{\includegraphics{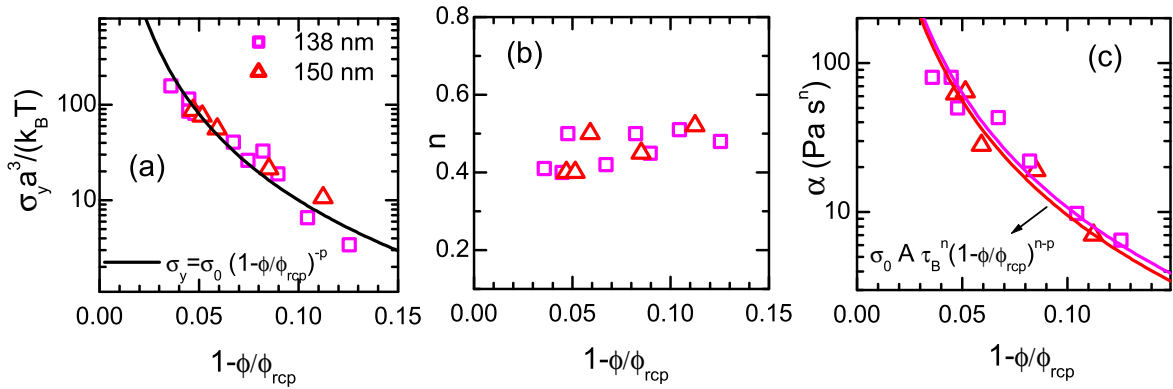}}
\caption{Bulk HB parameters for $a=138$~nm and $a=150~$nm versus $1-(\phi/\phi_{rcp})$. (a) Normalized yield stress. Line: $\sigma_y=\sigma_0[1-(\phi/\phi_{rcp})]^{-p}$ with $p=3$ and $\sigma_0=0.01 k_BT/a^3$. (b) the HB epxonent $n$. (c) The HB parameter $\alpha$. Lines: $\alpha=A\sigma_0 \tau_B^n [1-(\phi/\phi_{rcp})]^{n-p}$ with $A=10$, $n=0.45$, $p=3$ and $\tau_B=0.0285~$s and $\tau_B=0.036~$s the Brownian times.}
\label{fig_bulkrheovsphi}
\end{figure}
with $\tau_m=\tau_B(1-\phi/\phi_{rcp})$ a characteristic microscopic $\phi$ dependent (in-cage relaxation) time scale in the glass, $\tau_B$ the Brownian time and $A$ given by $A=\alpha/[\sigma_y \tau_B^n(1-\phi/\phi_{rcp})^n]	$. While this phenomenological behavior has no solid theoretical basis, it is physically plausible and has been used succesfully in [\cite{Besseling2010}]. We do not further pursue its interpretation here, but instead turn to the behavior of the slip parameters.

\label{Sec_slipparam}

\subsection{$\protect\phi$ dependence for non-stick walls}

As mentioned, for smooth, non-stick walls the Bingham slip parameters $\beta$ and $\sigma_s$ in Eq.~(\ref%
{eq_slipstress}) show a strong increase with $\phi$. We show
the parameters extracted from the rheology in Fig.~\ref{fig_numphidep}%
(a),(b), where in (a) the lubrication parameter has been normalized by the
solvent viscosity and particle size, while in (b) the stress is normalized
by $k_BT/a^3$. We observe a characteristic increase of the parameters for $\phi/\phi_{rcp} \rightarrow 0$ 
for the different particle sizes, except for deviations for the $a=652~$%
nm particles \footnote{The latter might be partly due to a somewhat reduced sediment volume fraction $\phi_{sed}<\phi_{rcp}$ from centrifugation of this batch (as discussed in Sect.~\ref{subsec_colloids}), thus slightly overestimating of $\phi/\phi_{rcp}$.}.

\begin{figure}
\scalebox{0.8}{\includegraphics{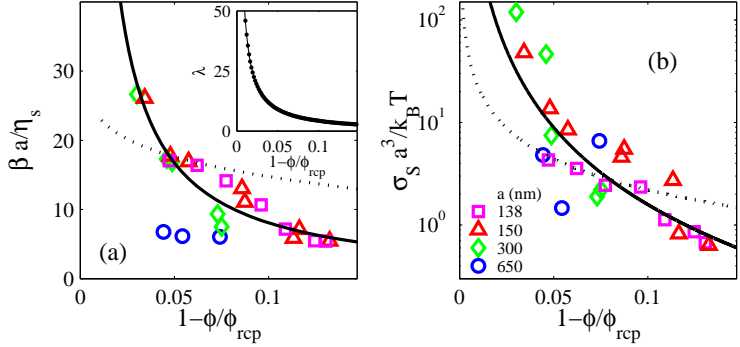}}
\caption{(a) Normalized lubrication parameter $\protect\beta a/\protect\eta%
_s $ versus $1-\protect\phi/\protect\phi_{rcp}$. Symbols: data for different
particle size (see (b)); full line: result based on Eq.~(\protect\ref%
{eq_empiricbeta}). Dotted line: result based on Eq.~(\protect\ref%
{eq:slipexpr}) and the explanation in the text. Inset: ($\bullet$) $\protect%
\lambda$ versus $1-\protect\phi/\protect\phi_{rcp}$ from a numerical
evaluation of Eq.~(\protect\ref{eq:lambda}). The full line represent $%
\protect\lambda \propto \Pi/\Pi_0$. (b) Normalized slip stress $\protect%
\sigma_s a^3/k_BT$ versus $1-\protect\phi/\protect\phi_{rcp}$, the dotted
line is $\protect\sigma_s = 0.45 \Pi$, the full line represents Eq.~(\protect
\ref{eq_empiricsigmas}) with $A=0.005$ and $m=2.5$.}
\label{fig_numphidep}
\end{figure}

We now attempt to rationalize this behavior in terms of the physical
properties of the suspension and the (hard) interaction with the wall. We
first discuss the lubrication parameter $\beta$. The linear increase of
stress with velocity, Eq.~(\ref{eq_slipstress}), implies that $\beta$ is
governed by hydrodynamic friction between the first layer of particles and
the wall, with an effective lubrication layer thickness independent of
velocity. For a single particle of radius $a$, centered at a distance $%
a+\delta$ from the wall, moving with constant velocity $v$ in a solvent of
viscosity $\eta_s$, the drag force it experiences is $\eta_s v a
f(\delta)$, where $f(\delta/a) \simeq 18.1-10 \ln(\delta/a)$ for $\delta \ll
a$. $f(\delta)$ reflects the wall induced hydrodynamic reduction of the
particle mobility [\cite{Goldman1967}] for no-slip boundary conditions.
For a distribution of particles with concentration $n(\delta)$ moving with
respect to the wall, the mean stress on the wall follows from integrating
over $\delta$. The lubrication parameter $\beta=\sigma/ v$ may thus be
written in normalized form as: 
\begin{equation}
\beta a / \eta_s = \sigma a/ \eta_s v = \int_0^{\infty} a^2 f(\delta )
n(\delta) d\delta.  \label{eq:slipexpr}
\end{equation}
We assume that beyond the first layer the fluid and colloids move together,
hence the integral can be cut off at $\delta=a$. With the further assumption
that solvent flow in the lubrication layer does not change the distribution $%
n(\delta)$ and that colloid interactions can be ignored, $\beta$ may be
evaluated from the ``equilibrium'' distribution $n_e(\delta)$. It was
demonstrated in [\cite{Henderson1984}] that the contact value $n_e(0)$
follows $n_e(0)=\Pi/k_BT$ with $\Pi$ the osmotic pressure of the suspension
at rest. While $\Pi(\phi)$ is uncertain for $\phi > \phi_g$ ([\cite%
{hermesdijkstra_epl10},\cite{Phan1996}, \cite{Tokuyama2007}]), a widely used
form is [\cite{Brady1995}]: 
\begin{equation}
\Pi = 2.9\Pi_0/(1-\phi/\phi_{\mathrm{rcp}}), ~~~~~~\mathrm{with}~~~~~\Pi_0 =
3 \phi k_B T/(4\pi a^3).
\end{equation}
Simulations for $\phi<0.5$ in [\cite{Henderson1984}] further showed that in
the first layer $n_e$ decreases as $n(\delta)=n_e(0) \exp(-3\lambda
\delta(1+\delta/a+\delta^2/3a^2)/a)$, with $\lambda$ a $\phi$-dependent
parameter. Employing this form also for the colloidal glass, $\lambda(\phi)$
can be extracted by equating $\phi$ in the first layer to the bulk value: 
\begin{equation}
\frac{2\pi a^2}{3}\int_0^a n(\delta) d\delta=\phi \sim a^2\int
n_e(0)\exp(-\lambda \delta/a) d\delta \sim n_e(0)a^3/\lambda
\label{eq:lambda}
\end{equation}
Numerical solution of the l.h.s. of Eq.~\ref{eq:lambda} gives $\lambda(\phi)
\propto \Pi(\phi)/(k_B T/a^3)$ (inset to Fig.~\ref{fig_numphidep}(a)), as
also confirmed by the scaling in the r.h.s. of Eq.~\ref{eq:lambda}. Using $%
\lambda(\phi)$ we then calculate $\beta$ from Eq.~\ref{eq:slipexpr}. The
result, shown by the dashed line in Fig.~\ref{fig_numphidep}(a), is finite
for all $\phi$ and diverges as $\phi \rightarrow \phi_{rcp}$, approximately
as $\beta a /\eta_s \propto - \ln(1-\phi/\phi_{rcp})$. While this qualitatively
accounts for the data, and Eq.~\ref{eq:slipexpr} matches the experimental $\beta\propto \eta a^{-1}$ scaling
(Figs.~\ref{Tdep} and \ref{fig_numphidep}(a)), the predicted $\beta(\phi)$ clearly 
does not properly describe the experimental results.

A possible explanation for the discrepancy is that the chosen form for $%
f(\delta)$ only applies for a single particle. However, while colloid
interactions are known to limit the wall induced reduction of diffusion in
concentrated colloidal liquids [\cite{Michailidou2009}], recent simulations [%
\cite{Swan2010}] show that for small $\delta$ the logarithmic $f(\delta)$
nevertheless holds. An alternative explanation is that the ``equilibrium''
form $n_e$ does not correctly represent the near wall particle distribution.
This can be due to the fact that already at rest the structure is out of
equilibrium (since $\phi > \phi_g$), possibly combined with the layering or
a change in concentration in the wall layer as observed in [\cite{Dullens}].
Equally likely is that $n(\delta)$ differs from $n_e$ due to the actual flow present
in the slip layer, similar to dilute systems [\cite{Polverari1995}]. This
non-equilibrium effect can be quantified by comparing the shear rate in the
lubrication layer with the inverse timescale $1/\tau_m$ for ``cage''
exploration (discussed at the start of this section), via a ``wall'' P\'{e}clet number $\mathrm{Pe}_w=\tau_m v_s /
\xi = \tau_m (\sigma-\sigma_s)/\eta_s$ ($\xi$ is a mean layer thickness
discussed below). Here $\tau_m$ is estimated using the mean free particle
space (also described below) and the short time diffusion coefficient [\cite%
{BradyCOColIntSci96_rheodifoverview}] as $\tau_m =
3\tau_B[1-\phi/\phi_{rcp}]$, in line with $\tau_m$ inferred from the $\phi$-scaling of the bulk rheology at the start of this section. Using the Brownian time $\tau_B \gtrsim 30~$ms
and $\sigma-\sigma_s \gtrsim 0.2~$Pa, we have $\mathrm{Pe}_w = O(1)$ and
larger for our data in the slip regime. Thus, during slip $n(\delta)$ is
indeed expected to differ considerably from $n_e$ [\cite{BradyJCP93}], but
further measurements are required to confirm this.

Empirically, the bulk of the experimental data are well described by the
form: 
\begin{equation}
\beta a /\eta_s \equiv a/\xi \simeq 0.9/(1-\phi/\phi_{rcp}) \simeq
\Pi/3.2\Pi_0,  \label{eq_empiricbeta}
\end{equation}
where the l.h.s. defines the mean lubrication layer thickness $\xi \propto 1/\Pi$. In fact, the mean spacing $\langle
s \rangle$ between colloids in the bulk is estimated as $\langle s \rangle
=a((\phi_{rcp}/\phi)^{1/3}-1)\simeq (a/3)(1-(\phi/\phi_{rcp}))$, hence the
empirical form implies $\xi \simeq 3.4 \langle s \rangle$. Using $n_e$ to
calculate $\xi$ gives similarly $\xi = a^2 \int_0^a \delta~ n(\delta) d
\delta \simeq 2 a \Pi_0/\Pi \simeq 2 \langle s \rangle$, but this form can
also result from other distributions, different from the 'equilibrium' one.
Overall, the analysis suggests that the non-equilibrium behavior in the
lubrication layer underlies the behavior of $\beta(\phi)$. Note
that our $\beta(\phi)$ matches the predicted $\phi$ dependence of the high
frequency viscosity or inverse short time diffusion constant in concentrated
suspensions [\cite{BradyCOColIntSci96_rheodifoverview}], for which, to our
knowledge, no experimental verification yet exist for $\phi \rightarrow
\phi_{rcp}$.

Next we discuss the behavior of the slip threshold $\sigma_s$. As seen in
Fig.~\ref{fig_numphidep}(b), $\sigma_s \propto 1/a^3$, suggesting a relation
to the osmotic pressure. In this context, $\sigma_s$ and $\Pi$ may be
naively linked via a phenomenological Coulomb friction mechanism using $%
\sigma_s = \mu \Pi$ with $\mu$ the friction coefficient. Comparing this with
the data, we see that a value $\mu \simeq 0.45$ can describe $\sigma_s$ at
intermediate $1-\phi/\phi_{rcp}$. Furthermore, recalling the experiments
with two uncoated surfaces in Fig.~\ref{fig_op2slip}, we found that the
ratio between $\sigma_s$ at the bottom plate and the cone is insensitive to $%
\phi$, Fig.~\ref{fig_op2slip}(b), which might support the Coulomb friction
scenario. However, the overall $\phi$ dependence of the data in Fig.~\ref%
{fig_numphidep}(b) is inconsistent with $\sigma_s \propto \Pi$ for constant $%
\mu$. The data are considerably more scattered than for $\beta(\phi)$ and 
also exhibit unsystematic variation of $\sigma_s$ within a factor $2$ for
different glass cleaning methods. While preventing a precise description
of the divergence, the data are reasonably described by: 
\begin{equation}
\sigma_s a^3 /k_B T \simeq A (1-\phi/\phi_{rcp})^{-m},
\label{eq_empiricsigmas}
\end{equation}
with $A \simeq 0.005$ and $m \simeq 2.5$. This dependence mimics the
behavior of the yield stress, $\sigma_y a^3/k_B T \simeq
0.01 (1-\phi/\phi_{rcp})^{-3}$. Thus, $\sigma_s/\sigma_y=O(0.1)$ and decreases weakly with $\phi$ as $%
\sigma_s/\sigma_y \simeq 0.5(1-\phi/\phi_{rcp})^{0.5}$. Therefore $
\sigma_s(\phi)$ may represent the stress required for the presumed
reorganization of the particle distribution when slip sets in and the
similarity between $\sigma_s(\phi)$ and $\sigma_y(\phi)$ could be connected
to a similar change of the particle distribution due to 'cage' breaking for $%
\sigma=\sigma_y$.

\begin{figure}
\scalebox{1}{\includegraphics{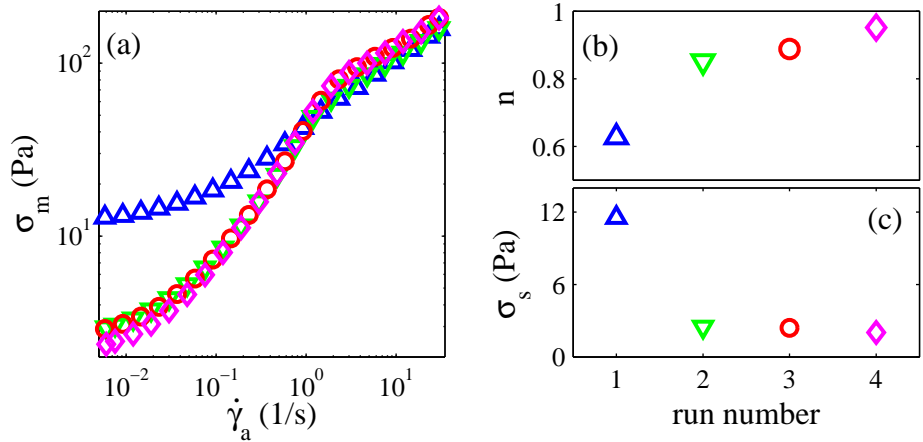}}
\caption{(a) Measured stress $\protect\sigma_m$ versus $\dot\protect\gamma_a$
for a suspension with $a=138$~nm, $\protect\phi=0.61$ in RI-matching
solvent, immediately after loading, going from small to large $\dot\gamma_a$ ($\triangle $), followed by large to small $\dot\gamma_a$ ($\bigtriangledown$%
), again small to large($\circ $) and finally large to small $\dot\gamma_a$ ($%
\diamond $). (b) The power $m$ and (c) $\protect\sigma_s$, both obtained from a fit of the low $\dot\protect%
\gamma_a$ branch to $\protect\sigma_m-\protect\sigma_s \propto \dot\protect%
\gamma_a^m$, versus run number.}
\label{fig_disc}
\end{figure}

A definite interpretation of $\sigma_s$ is thus still lacking. We conclude
by showing that nevertheless normal stresses in the system do seem to affect 
$\sigma_s$. For concentrated samples, the rheology of the system may show a
transient behavior after initial sample loading, associated with local shear
thickening behavior during loading. This is shown in Fig.~\ref{fig_disc}
where a sample is submitted to repeated low-to-high-to-low shear rate cycles
after loading. Fitting the low shear part of the first flow curve (%
\Blue{$\triangle$}) to $\sigma_m-\sigma_s \propto \dot\gamma_a^m$, we find $%
m \simeq 0.65$, i.e. in between the exponent $1$ for Bingham slip and $0.5$
for bulk flow, indicating a mixture of shear and slip. For
the first run $\sigma_s$ is significantly larger than for the next cycle(s),
where both $\sigma_s$ and $m$ reach a constant value $\sigma_s\simeq 2$~Pa
and $m\simeq 1$. Assuming that local shear thickened regions lead to
relatively large local normal stresses, increasing $\sigma_s$, the reduced $%
\sigma_s$ in repeat runs is consistent with flow induced relaxation of these
local normal stresses.

\subsection{Effect of wall interaction}

\label{sec_wallinteraction}

\begin{figure}[tbp]
\center\scalebox{0.5}{\includegraphics{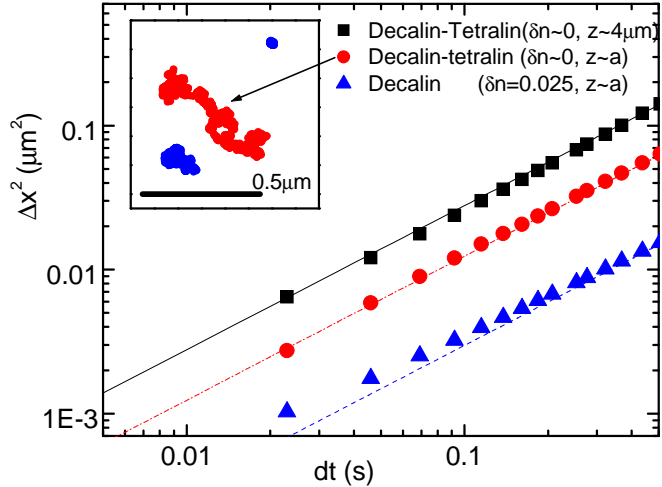}}
\caption{Mean squared displacement of $a=652~$nm particles versus time in
dilute suspensions in a decalin-tetralin mixture far from the glass ( 
\Black{$\blacksquare$}), close to the glass ($\bullet$) and close to the
glass in decalin ($\blacktriangle$). The lines represent diffusive behavior $%
\Delta x^2=2 D t$ with $D=0.139~\protect\mu$m$^2/$s, $D=0.062~\protect\mu$m$%
^2/$s and $D=0.014~\protect\mu$m$^2/$s from top to bottom. Inset:
trajectories at $z\sim a$ of particles in decalin (the stuck particle and
short trajectory) and a single long particle trajectory in decalin-tetralin (%
$\bullet$).}
\label{fig_diff}
\end{figure}

As discussed, the slip response for HS glasses strongly depends on the colloid-wall vdW
interaction \footnote{Note that, generally, attractive colloid-wall forces alone do not guarantee elimination of slip. For e.g. flocculated gels, with a heterogeneous network of aggregates, a smooth, attractive (or even particle coated) wall can act as an easy `fracture' plane resulting effectively in slip [\cite{Buscall1993}]. In these cases larger scale wall roughness - of the order of the aggregate size - is needed to prevent slip.}. As shown in Fig.~\ref%
{fig_im_nim}, for sufficient attraction slip vanishes, implying that the
slip threshold stress $\sigma _{s}>\sigma _{y}$. To characterize directly
the attraction, we imaged dilute suspensions of the $a=652$~nm particles and
analyzed the near-wall motion both in RI-matching and RI-mismatching
solvents, without flow. In Fig.~\ref{fig_diff} we show the mean squared
displacements (MSD) for the two cases, for particles imaged at the surface
(i.e. $z\simeq a$ within the microscope $z-$ resolution) and the MSD away
from the surface. As expected, [\cite{Goldman1967, Sharma2008,
Michailidou2009}] the MSD is smaller close to the wall than in the bulk.
Moreover, with RI-matching the near wall MSD is five times larger than in
the RI-mismatching solvent; in the latter case particles are (temporarily) stuck to the surface, evidencing the vdW attractions. With
RI-matching the mobility is enhanced, vdW forces are reduced and
insufficient to stick particles to the surface.

\begin{figure}
\scalebox{1}{\includegraphics{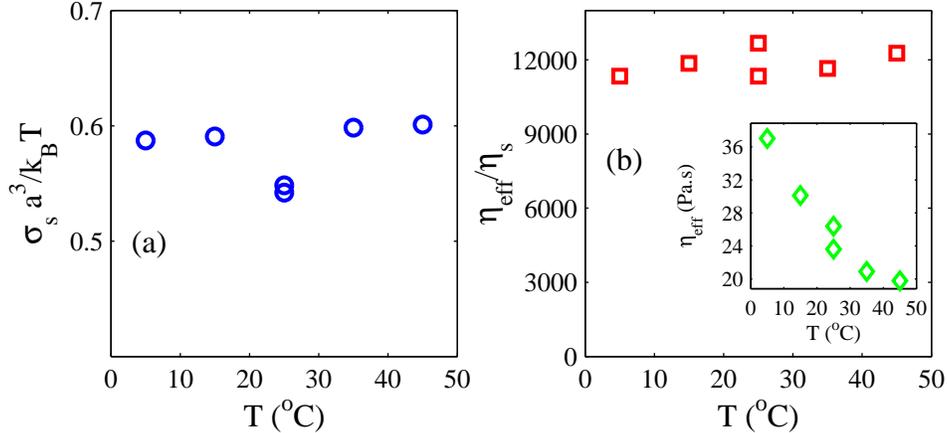}}
\caption{Temperature dependence of (a) the normalized Bingham slip stress
and (b) Bingham viscosity $\protect\eta_{\mathrm{eff}}(T)$ normalized by the
temperature dependent solvent viscosity $\protect\eta_{\mathrm{s}}(T)$. Data
are for $\protect\phi=0.59$, $a=138$~nm using a coated cone and smooth
glass. The inset to (b) shows the un-renormalized data $\protect\eta_{%
\mathrm{eff}}(T)$.}
\label{Tdep}
\end{figure}

Even though the vdW interaction is strongly reduced by RI-matching, it
cannot be completely suppressed. To study the role of remaining vdW forces
on the slip stress $\sigma _{s}$ (see Eq.~\ref{eq:binghameta}), we have
measured flow curves on the same sample, with smooth glass and coated cone,
at various temperatures $T$. Changing $T$ changes the RIs (mainly of the
solvent) and might thus be observable in the dependence of $\sigma _{s}$ on $%
T$. For all temperatures measured ($5-45$~C$^{\circ }$) the flow curves
exhibited the Bingham slip branch (data not shown), from which we extracted $%
\sigma _{s}(T)$ and the effective viscosity $\eta _{\mathrm{eff}}(T)$ via
Eq.~\ref{eq:binghameta}. Figure~\ref{Tdep}(a) shows that the normalized slip
stress, $\sigma _{s}a^{3}/k_{B}T$, is nearly temperature independent, but
exhibits a small drop of $\sim 10~\%$ for $T=25^{o}$C. For decalin $\partial
_{T}n_R=-4.4\cdot 10^{-4}$~K$^{-1}$, introducing a temperature dependent
solvent refractive index $n_R$ in the calculations presented in App.~\ref{B1} can be used to obtain a $T$-dependent particle-wall
interaction. For
what interests us here, there is a temperature interval where $n_{\mathrm{
PMMA}}<n_{\mathrm{R,solvent}}<n_{\mathrm{R,glass}}$ and thus colloid-wall
interactions may become slightly repulsive (see Table 3 below). The observed
minimum of $\sigma _{s}$ can qualitatively be associated with such slight
repulsion, although a calculation suggests that this should occur
around 40$^{\circ }$C rather than 25$^{\circ }$C as observed experimentally
(figure \ref{Tdep}(a)). We attribute the difference to the approximations
made in the calculation of the vdW interactions (App.~\ref{B1}). Overall,
we conclude that vdW forces do affect $\sigma_{s}$ but, in the range of
RI-mismatch considered here, have only a modest effect. The data for
the Bingham viscosity $\eta _{\mathrm{eff}}(T)$, inset to Fig.~\ref{Tdep}%
(b), also provide useful information. While $\eta _{\mathrm{eff}}(T)$
exhibits a clear temperature variation, when normalized by the temperature
dependent solvent viscosity, $\eta _{\mathrm{s}}(T)$ no $T$ dependence is
detected (Fig.~\ref{Tdep}(b)) suggesting that the viscous slip is due to a
lubrication layer of pure solvent between the colloids and the wall.

Rheology experiments along with near-wall motion measurements evidenced that sufficiently strong vdW attraction between colloids and walls
suppresses slip in the HS glasses even with smooth walls. Changing the index matching influences both the vdW interaction between two particles and the particle-wall interaction. We calculated the interaction $V_{pw}$ between
the wall and a colloid of radius $a$ separated by a distance $%
\delta$, and the interaction energy $V_{pp}$ between two colloids separated by a distance $2\delta $ using the formulas in
App.~\ref{B1}. The parameters for each solvent are given in
table ~\ref{tab_ind}. The resulting particle-particle and particle-wall
interactions for the various $a$ and the different solvents are shown in
table~\ref{tab3} using $\delta =10$~nm as thickness of the steric
layer. In decalin, particle-particle vdW attractions are weaker than $%
k_{B}T/10$ and thus can be neglected. However, the particle-wall (glass) vdW
attraction is of the order of $k_{B}T$ for the larger particles explaining
the significant tendency of these particles to stick to the surface,
although such attractions are clearly weaker than $k_{B}T/10$ for smaller
particles. Finally, the vdW attraction between colloids and the metallic
cone is stronger than $k_{B}T$ for all particle sizes suggesting that a
layer of stuck particles at the cone should be expected. In decalin-tetralin
mixture all attractions are at least an order of magnitude smaller than $%
k_{B}T$. Hence in index matching solvent particle-particle and particle-wall
interaction are reduced to their hard sphere counterparts.

\begin{table}[h!]
\begin{center}
\begin{tabular}{|c|c|c|c|c|c|}
\hline
Component & decalin & decaline-tetraline & PMMA & swollen PMMA & glass \\ 
\hline
$n_R$ & 1.47 & 1.51 & 1.49 & 1.51 & 1.523 \\ \hline
$\epsilon$ & 2.43 & 2.63 & 2.6 & 2.6 & 3.4 \\ \hline
\end{tabular}%
\end{center}
\caption{Dielectric permittivity $\protect\epsilon $ and index of refraction 
$n_R$\ for the solvents, glass, and PMMA.}
\label{tab_ind}
\end{table}

Still, even with a good index matching, some residual van der Waals forces
are present (see table~\ref{tab3}). Moreover, in decalin-tetralin mixtures,
the particle-glass surface interaction is positive, which denotes a repulsive
force. This is due to the fact that in this case $\epsilon _{1}<\epsilon
_{3}<\epsilon _{2}$ (see table~\ref{tab_ind}). Even if such a repulsive
interaction should enhance slip, the small energies involved allow us to
mostly neglect this effect. Two other factors that can affect the slip
parameter by modifying the friction coefficient are the variations of the
index-matching solvent composition and the polymer stabilizing layer between
different batches and particles, and the glass plates that are replaced in each
measurement. These factors may introduce some experimental uncertainty
resulting in a variation of $\sigma _{s}$ by up to a factor of 2 between two
different experiments. Thus, the friction coefficient may vary both for
different particle sizes and between different experiments with the same
sample.

\begin{table}[h!]
\begin{center}
\begin{tabular}{|c|c|c|c|c|}
\hline
sample name & asm340 & asm247 & asm209 & asm195 \\ \hline
$a$ (nm) & 138 & 150 & 300 & 652 \\ \hline
$V_{pp}^{dec}$ ($k_{B}T$) & -0.011 & -0.012 & -0.032 & -0.082 \\ \hline
$V_{pw}^{dec}$ ($k_{B}T$) & -0.099 & -0.11 & -0.24 & -0.55 \\ \hline
$V_{pp}^{mix}$ ($k_{B}T$) & -67 e-7 & -76e-7 & -20 e-6 & -51 e-6 \\ \hline
$V_{pw}^{mix}$ ($k_{B}T$) & 0.00097 & 0.0011 & 0.0024 & 0.0054 \\ \hline
$V_{p-steel}^{dec}$ ($k_{B}T$) & -4.58 & -5.05 & -11.1 & -25.4 \\ \hline
$V_{p-steel}^{mix}$ ($k_{B}T$) & 0.0076 & 0.0084 & 0.018 & 0.042 \\ \hline
\end{tabular}%
\end{center}
\caption{van der Waals interactions for different particle sizes $a$, surfaces
and solvents: particle-particle interaction in decalin $V_{pp}^{dec}$,
particle-wall interaction in decalin $V_{pw}^{dec}$, particle-particle
interaction in decalin-tetralin $V_{pp}^{mix}$, particle-wall
interaction in decalin-tetralin  $V_{pw}^{mix}$, as well as the particle-steel cone interaction for decalin ($V_{p-steel}^{dec}$) and for decalin-tetralin ($V_{p-steel}^{mix}$%
).}
\label{tab3}
\end{table}

\section{Shear localization}

\label{sec_localization}

So far we have focused on results for which, at a given $r$ (i.e. gap size)
the shear rate $\dot{\gamma}(z)$ is essentially uniform over the gap when
the suspensions start to yield. However, in experiments where slip is suppressed by coating both
cone and plate, we have observed shear
banding for small shear rates (near $\sigma _{y}$) and large $\phi $. An
example is shown in Fig.~\ref{fig_localisation}(a). For large rates, $v(z)$
is approximately linear, but for $\dot{\gamma}\leq 3~$s$^{-1}$ the profiles
become strongly nonlinear, with shear localization detected near the walls
and a vanishing shear rate in the bulk for the smallest $\gamma_a$. In [%
\cite{Besseling2010}], we have shown that this behavior, with a \textit{%
continuous} variation of $\dot{\gamma}(z)$ over the gap, can be explained by
very small concentration gradients ($\delta \phi \lesssim 0.003$), caused by
a dilation-like flow instability due to shear concentration coupling (SCC) [%
\cite{Schmitt1995}]. This is qualitatively different from other soft glasses
[\cite{Besseling2010}] and cannot be explained using earlier models for
heterogeneous glassy flow, e.g. involving specific wall rheology [\cite%
{Bocquet2009}]. This instability and the associated nonlinearity in $v(z)$
sets in below a typical rate $\dot{\gamma}_{c}(\phi )$ which becomes
appreciable only for large $\phi $.

While localization is most easily observed for coated walls, where the
average bulk shear rate $\langle\dot\gamma\rangle=[v(z_g)-v(z=0)]/z_g$
equals $\dot\gamma_a$, we have also detected nonlinear profiles for smooth
walls near the slip to yield transition, $\langle\dot\gamma\rangle \to 0$,
for large $\phi$, Fig.~\ref{fig_localisation}(b). For $%
\dot\gamma_a=3-5~$s$^{-1}$ the suspension has started to yield, i.e. $%
\langle\dot\gamma\rangle>0$, but the $\dot\gamma(z)$ is
strongly reduced (enhanced) near the smooth (rough) wall compared to $%
\langle\dot\gamma\rangle$. Indeed, for $\dot\gamma_a=5~$s$^{-1}$ the mean
shear rate is $\langle\dot\gamma\rangle \sim 0.2\dot\gamma_a=1~$s$^{-1}$,
similar to the value of $\dot\gamma_a$ below which banding becomes
significant for rough walls, Fig.~\ref{fig_localisation}(a). Due to such nonlinear flow near yielding, it is difficult to determine $\sigma_{y}$ very precisely for $\phi/\phi_{rcp}\gtrsim
0.94$. The slightly reduced concentration $\phi (z)<\phi $ in the fluidized
bands allows flow for $\sigma$ (very) slightly below the average $\sigma_y$. Thus
the determined $\sigma_{y}$ for large $\phi $ may be slightly
underestimated, preventing exact calculation of $\dot{\gamma}(r)$ for large $%
\phi $ where $\dot{\gamma}_{c}$ is large. Therefore, we have only analyzed
the slip to yield transition for $[v(z_{g})-v(z=0)]/z_{g}>\dot{\gamma}_{c}$,
i.e. where the induced bulk flow is essentially linear.

\begin{figure}[tbp]
\scalebox{1}{\includegraphics{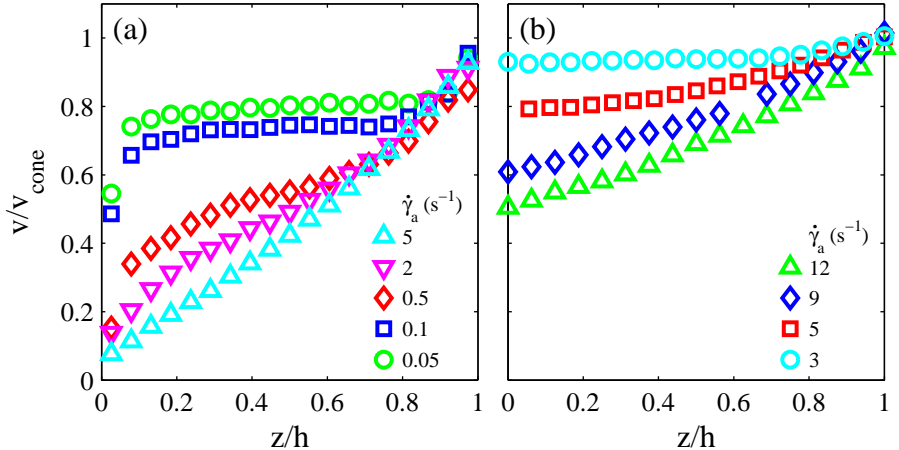}}
\caption{(a) Velocity profiles for $\protect\phi=0.62$, $r=5$~mm and both
surfaces coated, for various $\dot\protect\gamma_a$. (b) Same as (a) but for
coated cone, smooth plate and $r=5.5$~mm.}
\label{fig_localisation}
\end{figure}

\section{Slip below the glass transition}

\label{section_slip_liq}

In our earlier work~[\cite{Ballesta2008}], we reported that for
concentrations well below the glass transition, $\phi <\phi _{g}$, the flow
curves and velocity profiles showed no indication of slip or shear-banding
independent on the surface roughness. Figure \ref{fig_liquid} illustrates
this for $\phi =0.52$ where the flow curves for coated and smooth surfaces
are essentially identical and the flow profiles (Fig.\ref{fig_liquid}~(b))
are very close to linear for all $\dot{\gamma}_{a}$. However, recent
velocimetry data for suspensions closer to the glass transition clearly
reveal slip at the lowest applied shear rate, although the effect is not
detectable in the rheology of these samples.

\begin{figure}
\scalebox{1}{\includegraphics{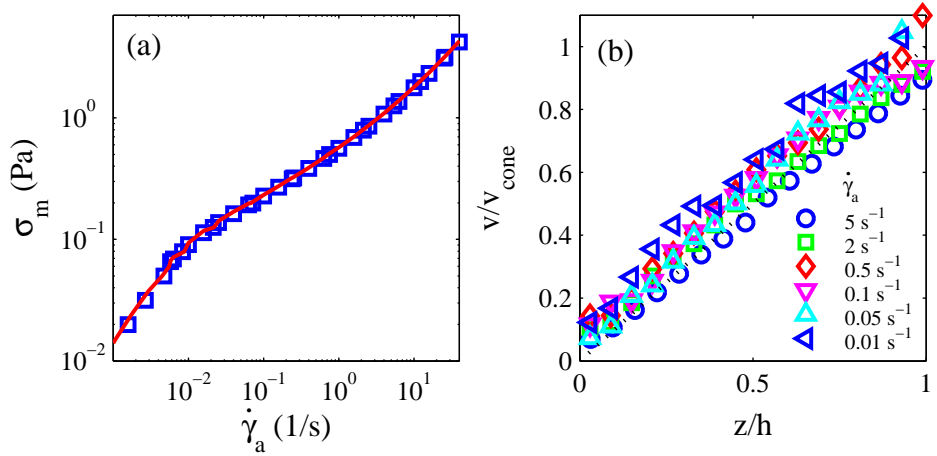}}
\caption{(a) Flow curve for an RI-matched suspension with $a=138$~nm, $%
\protect\phi=0.52$, with smooth ($\square$) and rough walls (full
line). (b) $v(z)$ for smooth walls at $r=5.5~$mm and for
various $\dot\gamma_a$.}
\label{fig_liquid}
\end{figure}

Figure~\ref{fig_liquidslip} shows the flow curve and velocity profiles for a 
$\phi =0.535$ suspension with coated cone and smooth glass surface. In the
flow curve, inset to Fig.~\ref{fig_liquidslip}(a), the low $\dot\gamma_a$ 
Newtonian behavior could not be resolved; only the shear thinning
behavior could be detected.
However, the velocity profiles in Fig.~\ref{fig_liquidslip}(a) clearly show
slip, but with a finite bulk shear rate $\dot\gamma>0$ (no plug flow),
similar to what is observed in glassy samples for $\sigma>\sigma_y$. We extracted the dependence of $\dot\gamma$ on $v_s$ for various gap sizes. The results in Fig.~\ref%
{fig_liquidslip}(b) show that $\dot{\gamma}=C
v_{s}^{2}$. This is consistent with the model presented earlier, but taking
into account the fact that in the concentrated liquid regime (here $\phi
=0.535$) the yield stress is absent. Using $%
\sigma _{s}=0$, we obtain $\sigma \simeq \beta v_{s}$ and $\sigma=\alpha \dot{%
\gamma}^{0.5}$ giving $C=(\beta/\alpha)^{2}$. From the imaging data in Fig.~\ref%
{fig_liquidslip}(b) we find $C=0.13 \cdot 10^{12}~$s/m$^2$ at this $\phi$. To compare this with the rheology, a fit of the flow curve gives $\alpha
=0.177$~Pa.s$^{0.5}$, while $\beta$ follows from extrapolation of the relation $\beta =0.9\eta
_{s}a^{-1}(\phi _{rcp}-\phi )^{-1}$ in the glassy state. This yields $\beta \simeq 5.8\cdot 10^{4}$~Pa.s$/$m at this $\phi$, giving $(\beta/\alpha)^2=0.11 \cdot 10^{12}~$s/m$^2$, in very reasonable agreement with the data. The model can thus also describe residual slip of liquids, assuming $\sigma_{s,y}=0$.

\begin{figure}
\scalebox{1}{\includegraphics{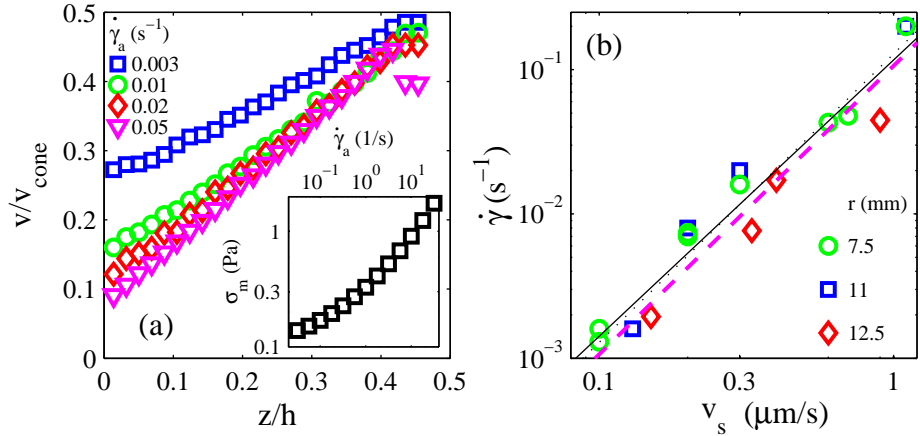}}
\caption{(a) Normalized flow profiles at $r=12.5~$mm for $\protect\phi=0.535$
at various $\dot\protect\gamma_a$ using a smooth cone and smooth glass.
Inset: measured flow curve. (b) Local shear rate $\dot\protect\gamma$ versus
slip velocity $v_s$ at various $r$. Full line: fit to a power law $\dot%
\protect\gamma\propto v_s^{\protect\nu}$ with $\protect\nu=1.92\pm0.3$.
Dotted line: $\dot\protect\gamma= C v_s^2$ with $C=0.13 \cdot 10^{12}~$s/m$%
^2 $. Magenta discontinued line: $\dot\protect\gamma= C v_s^2$ with $C=(%
\protect\beta/\protect\alpha)^2$ (see text).}
\label{fig_liquidslip}
\end{figure}

We can compare the slip behavior of glasses and liquids further as follows. The rheology of
liquid-like suspensions at low $\dot \gamma$ is characterized by a Newtonian flow ($%
\sigma =\eta _{0}\dot{\gamma}$ with $\eta
_{0}=\eta_0(\phi)$ the zero shear viscosity), which evolves towards nonlinear shear thinning behavior at higher shear rates ( $\sigma \propto \dot{\gamma}^{n}$). As a result of the low shear rate Newtonian behavior, any applied stress
results in shear at all $r^{\prime }$s for any such small $\dot{\gamma%
}_{a}$. At small $\dot\gamma$, the reduced
apparent viscosity is $\eta _{m}=\eta _{0}/(1+\eta _{0}/\beta h)$ (in plate-plate approximation\footnote{In cone-plate, setting $h=r \tan(\theta)$ and integrating as in Eq.~\ref{eq_stressint}, the refined expression for the apparent viscosity of a Newtonian liquid with slip is $\eta_m/\eta_0=1-2x_c[1-x_c\ln(1+x_c^{-1})]$ with $x_c=\eta_0/(\beta\theta r_c)$.}). A
rough estimate of the difference between slip and no slip measurements can be obtained by using the phenomenological 
form for the divergence $\eta
_{0}(\phi)$ on approaching the glass transition (e.g. [\cite{Meeker1997}]): $\eta _{0}=\eta
_{s}(1-\phi \phi _{g}^{-1})^{-2}$ and the earlier mentioned extrapolation of $\beta =0.9\eta
_{s}a^{-1}(\phi _{rcp}-\phi )^{-1}$ to $\phi< \phi_g$.
For $\phi =0.52$, $a=138$~nm and $h=50$~$\mu $m this leads to $\eta
_{m}=0.9588\eta _{0}$. A $\sim 4\%$ difference is roughly within the
experimental uncertainty, which explains why we did not note slip at low $\phi$ in [\cite%
{Ballesta2008}]. However, as $\phi $ approaches $\phi _{g}$, $\eta _{0}$
diverges while $\eta _{m}$ tends towards $\beta h$ and slip becomes apparent. For intermediate and large $\dot\gamma$, where the rheology can be approximated by $\sigma \propto \dot{\gamma}%
^{n}$ with $n\simeq 0.5$ (as stated above, and in agreement with a semi-empirical expression in
\cite{Krieger1959}]), a calculation of the flow curves with and without slip shows that in all cases the relative difference between $\eta _{m}$ and $%
\eta $ increases with $\phi$ and decreases as $\dot{\gamma}_{a}$ increases,
similarly to the glassy state. Since in the low-shear Newtonian regime the difference is already small (except very close to $\phi_g$), the presence of slip is even more difficult to detect in the nonlinear rheology of liquids at larger rate. However, for $\phi>\phi_g$ the presence of a yield stress and plug flow leads
to a stress difference $\sim \sigma _{y}$ regardless of the geometry, which explains why slip is easily detected in the rheology of HS glasses both for low and intermediate shear rates.

\section{Discussion and conclusions}
\label{disc_concl}

The rheology and velocimetry results and the modeling we presented clearly show that local particle-wall interactions, the character of the boundaries, the geometry and $\phi$ dependent nonlinear bulk rheology strongly affect the measured rheology of concentrated HS suspensions. In general for yield stress fluids, if
a flow curve exhibits a `kink' and a stress drop at low $\dot\gamma_a$ (with a power
law $\dot{\gamma}_{a}^{m}$) that is absent when rough surfaces are used, one can conclude that ($i$) the sample slips at one or both surfaces and ($ii$) locally the transmitted stress is proportional to $v_{s}^{m}$, independent of the geometry. The detailed phenomenological model we presented shows that, from a well characterized slip law $\sigma(v_s)$ and bulk rheology, the overall flow curve and local flow profiles can be accurately predicted (if more complex behavior such as shearbanding and possible non-stationary behavior can be ignored).

For our HS glasses with smooth non-stick walls, a velocity independent, but $\phi$-dependent, lubrication layer forms, leading to $m=1$, but vdW attractions, in non-RI-matched suspensions, easily suppress slip, leading to a slip stress $\sigma_s \geq \sigma_y$. Standard (non-imaging) rheology experiments for HSs are therefore unlikely to be affected by slip. Yet, with many recent studies of nonlinear colloidal flow focusing on microscopic properties [\cite{Cohen2006,Besseling2007}] via microscopy on RI matched suspensions, slip is an important ingredient, and we have shown here that the slip response can be related semi-quantitatively to a bulk property of the suspension (osmotic pressure). For other yield stress fluids, in particular jammed emulsions, the slip behavior may follow a different powerlaw, i.e. $m\simeq 0.5$ has been observed in [\cite{Meeker2004b,Seth2008}] due to elastohydrodynamic lubrication for non-repulsive smooth walls. {\it With} repulsion, $m\simeq 1$ is recovered, implying that our model of the slip-yield transition may carry over to emulsions. Further, emulsions with $\phi_g \lesssim \phi \lesssim \phi_{rcp}$ exhibit HS like (Brownian) glassy behavior [\cite{Gang1999}], for which we thus expect similar Bingham slip behavior as for the HSs. 

In non-Brownian suspensions [\cite%
{Jana1995,Soltani,Kalyon2005}], slip is also characterized by $m\simeq 1$, i.e. $\sigma \propto v_s$, i.e. a lubrication layer with a thickness independent of $v_s$ [\cite{Kalyon2005,Yilmazer1989}]. Here, contrary to colloids, no
slip stress is observed, as expected from the (near) absence of osmotic or wall-interaction effects. However, the detailed mechanism for slip in non-Brownian systems is still unclear as shown by the different phenomenological
relations found: $\delta /a\simeq 0.125$ for concentrated but Newtonian suspensions ([\cite{Jana1995}],$\phi$-independent) , $\delta
/a\simeq 0.06-0.15$ (for pastes of polydisperse spheres [\cite{Soltani}]), and $\delta/a \simeq 2/[1-(\phi/\phi _{rcp})]$ ([\cite{Kalyon2005}], for systems including polydisperse and non-spherical particles). Interestingly, the latter is similar to our Eq.~\ref{eq_empiricbeta} for the slip of colloidal glasses, but lacks a theoretical basis for non-Bronwian systems. Moreover, in the latter two cases, slip was measured for non-uniform stress, such that shear induced migration may affect the interpretation.

We can also compare the results with those for (depletion) flocculated colloidal gels in [\cite{Buscall1993}]. There, a linear slip response has also been observed, with $\delta$ decreasing from $\sim ~1\mu$m to $\sim 10~$nm from $\phi\sim 0.2$ to $\phi\sim 0.55$, without siginificant dependence on particle size or colloid attraction strength. Here the nature of the slip layer is likely determined by the $\phi$ dependent aggregate lengthscale and structure, rather than the particle size. It is worth noting again that in these systems, even significant colloid-wall attraction is generally unable to suppress slip; unless the wall roughness is very large, the boundary typically acts as a weak `fracture' plane and a slip response is induced. Insight in the nature of this behavior and a theoretical understanding are still lacking. 

In conclusion, we have shown the existence of Bingham-type slip response in colloidal HS glasses near smooth non-stick walls. A phenomenological model quantitatively accounts for the global rheology and local flow profiles. Slip in HS glasses is effectively caused by Brownian motion, creating a lubrication layer and slip response governed by the suspensions osmotic pressure, evidenced by the particle size dependence and divergence for $\phi \rightarrow \phi_{rcp}$ of the slip stress and slip viscosity. For HSs, slip is suppressed by colloidal scale wall roughness or sufficient vdW wall attraction. Slip can also occur in concentrated liquid-like suspensions, but is {\it partial} ($\dot\gamma_a > \dot\gamma \neq 0$) due to absence of a yield stress. This is also described within the phenomenological model, but has only limited effect on the bulk rheology. Our study of HSs and the study of [\cite{Meeker2004b,Seth2008}] for emulsions, together with future similar studies for other yield stress fluids, should provide improved predictability of yield stress fluid flows in industrial processing and applications.

\subsubsection{Acknowledgements}

We thank K.N. Pham, J. Arlt and N. Pham for advice and help with the
experiments, A.B. Schofield for particle synthesis and sizing and M.E.
Cates, A. Morozov and D. Marenduzzo for useful discussions. R.B. and W.P.
acknowledge funding through EP/D067650 and EP/D071070/1. L.I. was funded by the EU network
MRTN-CT-2003-504712. G. P. and P. B. acknowledge EU funding from ToK
\textquotedblleft Cosines\textquotedblright\ (MTCDCT-2005-029944), NMP Small
\textquotedblleft Nanodirect\textquotedblright\ (CPFP7- 213948-2) and NoE
"SoftComp".

\appendix{}

\section{Local and global rheology in cone-plate geometry}

\label{A1}

\subsection{Slip at one surface}

\label{appsec_slipone}

This is the case in most of our experiments. Slip occurs at the bottom plate
and we set $\sigma_1=\infty$ (i.e. $v_1=0$) and $\sigma_2<\sigma_y$. In a
cone-plate, the relative stress inhomogeneity over the gap is $\simeq
\theta^2$. This is negligible in our case and we take $\sigma$ uniform at a
given $r$. Using Eqs.~(\ref{eq_slipstress},\ref{eq_ratevslip}) with $%
h=r\tan(\theta)$ we have: 
\begin{equation}  \label{temp2}
\sigma(r)=\beta r \tan(\theta) [\dot\gamma_a-\dot\gamma(r)]+\sigma_2.
\end{equation}
The stress measured with the rheometer follows from Eq.~(\ref{eq_stressint})
in Sec.~\ref{Sec_model}. We define the critical applied shear rate $%
\dot\gamma_{a,c}=(\sigma_y-\sigma_2)/\beta r_c\tan(\theta)$ such that for $%
\dot\gamma_a<\dot\gamma_{a,c}$, the bulk shear rate vanishes over the entire
geometry (regime I). The measured stress is then given by Eq.~(\ref%
{eq_sigetafI}) in Sec.~\ref{sec_coneplate}. When $\dot\gamma_a>\dot%
\gamma_{a,c}$ (regime II), we define $r_y=(\sigma_y-\sigma_2)/(\beta
\tan(\theta) \dot\gamma_a)$ such that $\dot\gamma=0$ for $r\leq r_y$ and $%
\dot\gamma>0$ for $r > r_y$. The integral in Eq.~(\ref{eq_sigetafI})
consists of two parts: solid body rotation for $r<r_y$ and slip and shear
for $r>r_y$. Using Eq.~(\ref{eq_1u_dg_dga}) and Eq.~(\ref{eq_param1slip})
with $h=r\tan(\theta)$ and Eq.~(\ref{temp2}) gives the following result: 
\begin{eqnarray}  \label{eq_sigmamtot}
&\sigma_m^{\mathrm{II}}=\sigma_y-\frac{\Delta\sigma^3}{3\sigma_E^2}-\frac{%
\sigma_R^2}{\sigma_E} \\
&+\frac{\sigma_R^2}{\sigma_E}\left(\frac{3\Delta\sigma}{4\sigma_E} +\left(%
\frac{1}{2}-\frac{\Delta\sigma}{4\sigma_E}\right) \sqrt{1+\frac{4\sigma_E^2}{%
\sigma_R^2}\left(1-\frac{\Delta\sigma}{\sigma_E} \right)}\right)  \nonumber
\\
&+\left(\frac{\sigma_R^3}{4\sigma_E^2}-\frac{\Delta\sigma^2\sigma_R}{%
4\sigma_E^2} \right) \log \left(\frac{2\frac{\sigma_E}{\Delta\sigma}-1+\sqrt{%
\frac{\sigma_R^2 }{\Delta\sigma^2}+\frac{4\sigma_E}{\Delta\sigma}\left(\frac{%
\sigma_E}{\Delta\sigma}-1 \right)} }{1+\frac{\sigma_R}{\Delta\sigma}}
\right),  \nonumber
\end{eqnarray}
with $\sigma_E=\beta \tan(\theta) r_c \dot\gamma_a$, $\sigma_R =\alpha \sqrt{%
\dot\gamma_a}$, and $\Delta\sigma=\sigma_y-\sigma_2$.

We have assumed that in regime II shear occurs only in the $z$ direction, 
\textit{i.e} $\partial_r v=0$. But because the (partial) slip velocity
depends on $r$, in general $\partial_r v\neq 0$ during shear flow. This
radial velocity gradient can be calculated from $v=\dot\gamma z$, where $%
\dot\gamma$ is given by Eq.~(\ref{eq_1u_dg_dga}) with $h=r \tan(\theta)$,
with the result: 
\begin{equation}
\partial_r v=z\partial_r \dot\gamma=\frac{z}{r}\left(\frac{\Delta\sigma}{%
h\beta}-2\dot\gamma_0+\frac{2\dot\gamma_0+\left(2\dot\gamma_a-\frac{%
3\Delta\sigma}{h\beta}\right)}{\sqrt{1+\frac{2}{\dot\gamma_0}%
\left(\dot\gamma_a -\frac{\Delta\sigma}{h\beta} \right)}} \right).
\end{equation}
Hence, $\partial_r v$ tends to $0$ for $r\rightarrow \infty$ and $%
r\rightarrow r_y$. However, approaching the boundary of the region with 
solid body, $r\rightarrow r_y$, the relative contribution $%
(\partial_r v)/\dot\gamma$ grows as: 
\begin{equation}
\partial_r v/\dot\gamma \simeq 2z/(r-r_y) < 2r\tan(\theta)/(r-r_y).
\end{equation}
The shear rate in the vorticity direction may thus be important for $r \simeq r_y$, but is negligible for $(r/r_y)-1 \gg
2\tan(\theta)$. For our case ($\theta=1^{\circ}$), one can neglect $%
\partial_r v$ in practice, as confirmed by the agreement with the experiments, e.g. in Fig.~\ref{fig_localrdep}.

\subsection{Slip at both surfaces}

\label{appsec_sliptwo}

We now also allow slip at the cone with $\sigma_2\leq\sigma_1<\sigma_y$. We
first discuss solid body rotation in regime $\mathrm{I_b}$, \textit{i.e.} $%
\sigma_1<\sigma<\sigma_y$. The suspensions angular velocity $\omega_{\mathrm{%
bulk}}$ in this regime is determined as follows. The slip velocities at the
top and bottom plates are given by $v_1 =(\omega-\omega_{\mathrm{bulk}})r$
and $v_2=\omega_{\mathrm{bulk}}r$, respectively, with $\omega=\dot\gamma_a%
\tan(\theta)$. Since the total, integrated
stress $\sigma_m$ at the top ($\sigma_{top}=\sigma_1+2\beta(\omega-\omega_{%
\mathrm{bulk}})r_c/3$) and bottom ($\sigma_{bottom}=\sigma_2+2\beta\omega_{%
\mathrm{bulk}}r_c/3$) surfaces are equal, the solid body rotation velocity in regime $\mathrm{I_b}$ is: 
\begin{equation}  \label{eq_omega_Ib}
v_{bulk}=r\dot \gamma_a \tan(\theta)/2+3(r/r_c)(\sigma_1-\sigma_2)/4\beta.
\end{equation}
As a result, for slip at both surfaces, the local stress $\sigma(r)$ differs
between the cone and the plate, by an amount $\Delta\sigma(r)=(\sigma_1-%
\sigma_2)\left(1-\frac{3r}{2r_c}\right)$. The rate $\dot\gamma_a^*$ where the transition from slip at the
plate ($\mathrm{I_a}$) to slip at the plate and cone ($\mathrm{I_b}$) occurs, 
is determined by equating the measured stress due to slip at the plate and
that for slip at both surfaces, Eqs.~(\ref{eq_sigetafI},\ref%
{eq_sigetafIb}):  
\begin{equation}  \label{eq_omega_Iab}
\dot\gamma_a^*= 3(\sigma_1-\sigma_2)/2\beta r_c\tan(\theta).
\end{equation}

When $\dot\gamma_a$ increases and regime II is approached, the above
calculation shows that shear will first occur at the edge of the bottom
plate for $\dot\gamma_{a,e}=\frac{4\sigma_y-3\sigma_1-\sigma_2}{2\beta
r_c\tan(\theta)}$. The local stress difference between the top and bottom
plate makes a complete description of shear propagation into the cell very
difficult. However since $\sigma_1-\sigma_2$ is often small compared to $%
\sigma_y$ some approximations can be made. We assume that shear occurs at
the same position $r_y$ for the cone and the plate and that in the sheared
region of the sample ($r>r_y$) the stress $\sigma(z)$ is uniform. For
infinite parallel plates we already found that the flow with slip at both
surfaces is described in the same way as slip at one surface by replacing $%
\beta \rightarrow \beta/2$ and $\sigma_2 \rightarrow(\sigma_1+\sigma_2)/2$,
see Eqs.~(\ref{eq_param1slip},\ref{eq_param2slip}). The same change of
variables applied to the cone-plate leads to $r_y=(2\sigma_y-\sigma_1-%
\sigma_2)/\beta\omega$ and a transition from regime $\mathrm{I_b}$ to $%
\mathrm{II}$ at $\dot\gamma_{a,c}=\frac{2\sigma_y-\sigma_1-\sigma_2}{\beta
r_c\tan(\theta)}$, which, with the approximation $\sigma_1-\sigma_2\ll
\sigma_y$, is equivalent to $\dot\gamma_{a,e}$ above. We then again have
shear and slip for $r>r_y$ and solid body rotation for $r<r_y$. The velocity
for $r<r_y$ follows, as before, from the balance between total stress on the
bottom plate and cone: 
\begin{equation}
v_{r<r_y}=\frac{r \dot \gamma_a \tan(\theta)}{2}\left(1+\frac{3}{2}\frac{%
\sigma_1-\sigma_2}{2\sigma_y-\sigma_1-\sigma_2} \right).
\label{eq_rotvelunshear}
\end{equation}

For $\dot\gamma_a\geq \dot\gamma_{a,c}$, the measured stress is still given
by eq.~(\ref{eq_sigmamtot}) with the following changes: $\beta \rightarrow
\beta/2$ and $\sigma_2 \rightarrow (\sigma_1+\sigma_2)/2$.

\subsection{Slip length}

\label{subannexe_ls}

The relation between bulk and applied shear rate in regime II can also be
expressed via a slip length. For slip at the bottom plate only, we have: 
\begin{equation}  \label{eq_sliplength}
l_s=v_s/\dot\gamma=h (\dot\gamma_a-\dot\gamma)/\dot\gamma
\end{equation}
with $v_{s}$ the slip velocity, $h$ the gap and $\dot\gamma$ and $%
\dot\gamma_a$ the local and applied shear rate, respectively. The full
dependence $l_s(\dot\gamma_a)$ follows directly from the relation $%
\dot\gamma(\dot\gamma_a)$, Eq.~(\ref{eq_1u_dg_dga}) and Eq.~(\ref%
{eq_param1slip}) in Sect.~\ref{Sec_model}. For large shear rate,
Taylor expansion of Eq.~(\ref{eq_1u_dg_dga}) in $\dot\gamma_a^{-1/2}$ results in: 
\begin{equation}  \label{eq_sliplength3}
l_s=h \sqrt{\frac{2\dot\gamma_0}{\dot\gamma_a}}+O(\dot\gamma_a^{-1}).
\end{equation}
Figure~\ref{fig_slip_length} shows both the exact form for $l_s$ from
Eqs.~(\ref{eq_sliplength},\ref{eq_1u_dg_dga}) and the high shear
approximation, showing the divergence of $l_s$ for $\dot\gamma_a \downarrow
\dot\gamma_y$ and the asymptotic decrease $l_s \sim 1/\sqrt{\dot\gamma_a}$.

\begin{figure}[tbp]
\center\scalebox{1}{\includegraphics{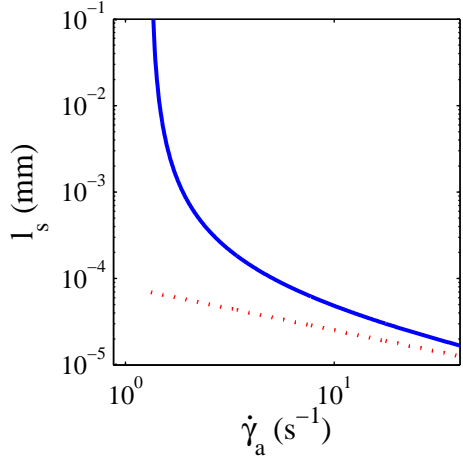}}
\caption{Slip length $l_s$ versus applied rate for $h=50~\protect\mu$m, $%
\protect\alpha=10~$Pa$\cdot$s$^{1/2}$, $\protect\beta=1.25\cdot 10^5~$Pa$\cdot$s$\cdot$m$^{-1}$, $%
\protect\sigma_y=10.6~$Pa and $\protect\sigma_s=2.39~$Pa. Full line is the
exact form using Eq.~(\protect\ref{eq_sliplength}), Eq.~(\protect\ref%
{eq_1u_dg_dga}) and Eq.~(\protect\ref{eq_param1slip}) in Sect.~\protect\ref%
{Sec_model}, dotted line represents Eq.~\protect\ref{eq_sliplength3}. }
\label{fig_slip_length}
\end{figure}

\section{van der Waals interactions}

\label{B1}

The van der Waals interactions between two particles and between a particle and a
wall were calculated according to [\cite{Gregory1981,Hunter_book}]. For a
particle of radius $a$ separated by a distance $\delta $ from the wall we
have: 
\begin{equation}
V_{pw}=-\frac{A_{132}a}{6\delta }\left( 1+\frac{\delta }{2a+\delta }+\frac{%
\delta }{a}\ln \left( \frac{\delta }{2a+\delta }\right) \right) ,
\end{equation}%
while the interaction between two colloids separated by a distance $%
2\delta $ is: 
\begin{equation}
V_{pp}=-\frac{A_{131}a}{24\delta }\left( \frac{2a}{2a+\delta }+\frac{%
2a\delta }{(a+\delta )^{2}}+\frac{4\delta }{a}\ln \left( \delta \frac{%
2a+\delta }{(a+\delta )^{2}}\right) \right) .
\end{equation}

For the interaction between the glass plate and particles, the Hamaker
constant $A_{132}$, where the indexes $i=1,2,3$ refer to PMMA, glass, and
solvent, respectively (table~\ref{tab_ind}), is approximately~[\cite{Lee2005}%
]: 
\begin{equation}
A_{132}=\frac{3k_{B}T}{4}~\prod_{i=1}^{2}\left( \frac{\epsilon _{i}-\epsilon
_{3}}{\epsilon _{i}+\epsilon _{3}}\right) +\frac{3h\nu _{e}}{8\sqrt{2}}~%
\frac{1}{\sum_{i=1}^{2}\sqrt{n_{i}^{2}+n_{3}^{2}}}\prod_{i=1}^{2}\left( 
\frac{n_{i}^{2}-n_{3}^{2}}{\sqrt{n_{i}^{2}+n_{3}^{2}}}\right) ,
\end{equation}%
while the Hamaker constant between the steel cone (denoted by subscript $2$) and the
particles ($1$) through solvent ($3$) is estimated using the approach of [%
\cite{Lipkin1997}]: 
\begin{equation}
A_{132}=\frac{3k_{B}T}{4}\frac{\epsilon _{1}-\epsilon _{3}}{{\epsilon
_{1}+\epsilon _{3}}}+3h\nu _{e}n_{1}(n_{1}-n_{3})\left( \frac{1}{2n_{1}}-%
\frac{1}{\sqrt{2(n_{1}^{2}+2+2\sqrt{2})}}\right) .  \label{eq_Ham2}
\end{equation}

Here $\epsilon _{i}$ is the dielectric permittivity, $n_{i}$ the respective
indexes of refraction, $h$ the Planck constant, and $\nu _{e}$ a
characteristic frequency which for simplicity is equated with the lowest energy adsorption
peak of PMMA in the ultraviolet $\nu _{e}=0.3 \times 10^{16}$~Hz (\cite{Hough1980}).

\end{document}